\documentclass[preprint]{elsarticle}
\usepackage{algorithm2e}
\usepackage[utf8]{inputenc}
\usepackage{graphicx
,amssymb,amsmath,tikz,tcolorbox,amsthm,lineno,url,fullpage}
\usepackage[]{algorithm2e}

\usepackage{url}

\usepackage{bm}

\newenvironment{claimproof}{\paragraph{Proof of the claim:}}{\hfill$\triangleleft$}

\newenvironment{proofsketch}{%
  \proof}{\endproof}

\date{}
\journal{Theoretical Computer Science}

\newtheorem{observation}{Observation}
\newtheorem{corollary}{Corollary}

\newtheorem{lemma}{Lemma}
\newtheorem{theorem}{Theorem}

\usepackage{thmtools}
\usepackage{thm-restate}

\usepackage{todonotes}

\usepackage{hyperref}
\usepackage{xcolor}
\hypersetup{
    colorlinks=true,
    linkcolor=blue,
    urlcolor=blue,
    citecolor=blue
}
\usepackage{cleveref}

\DeclareMathOperator{\rad}{rad}

\DeclareMathOperator{\MP}{mp}

\begin{document}
\begin{frontmatter}

\title{Multipacking and broadcast domination on cactus graphs and its impact on hyperbolic graphs\tnoteref{label1,label2}}
\tnotetext[label1]{A preliminary version of this paper entitled ``Multipacking and broadcast domination on cactus graphs and its impact on hyperbolic graphs'' has appeared in the proceedings of the {\it International Conference and Workshops on Algorithms and Computation} (WALCOM), pp 111–126, 2025 (\url{https://doi.org/10.1007/978-981-96-2845-2_8}). }

\author[1]{Sandip Das}
\ead{sandip.das.69@gmail.com}

\author[1]{Sk Samim Islam}
\ead{samimislam08@gmail.com}

\address[1]{Indian Statistical Institute, Kolkata, India}

\begin{abstract}
       
For a graph $G$, $ \MP(G) $ is the multipacking number, and $\gamma_b(G)$ is the broadcast domination number. It is known that $\MP(G)\leq \gamma_b(G)$ and  $\gamma_b(G)\leq 2\MP(G)+3$ for any graph $G$, and it was shown that $\gamma_b(G)-\MP(G)$ can be arbitrarily large for connected graphs. It is conjectured that  $\gamma_b(G)\leq 2\MP(G)$ for any general graph $G$. 
  We show that, for any cactus  graph $G$,  $\gamma_b(G)\leq \frac{3}{2}\MP(G)+\frac{11}{2}$. Although cactus graphs form a narrow graph class, we used some non-trivial techniques to provide the bound. These techniques make an important step towards generating a tighter bound for general graphs. We also show that $\gamma_b(G)-\MP(G)$ can be arbitrarily large for cactus graphs and asteroidal triple-free graphs by constructing an infinite family of cactus graphs which are also asteroidal triple-free graphs such that the ratio $\gamma_b(G)/\MP(G)=4/3$, with $\MP(G)$ arbitrarily large.  Moreover, we provide an $O(n)$-time algorithm to construct a multipacking of cactus graph $G$ of size at least $ \frac{2}{3}\MP(G)-\frac{11}{3} $,  where $n$ is the number of vertices of the graph $G$. The hyperbolicity of the cactus graph class is unbounded. For $0$-hyperbolic graphs, $\MP(G)=\gamma_b(G)$. Moreover, $\MP(G)=\gamma_b(G)$ holds for the strongly chordal graphs which is a subclass of $\frac{1}{2}$-hyperbolic graphs. Now it's a natural question: what is the minimum value of $\delta$, for which we can say that the difference $ \gamma_{b}(G) -  \MP(G) $ can be arbitrarily large for $\delta$-hyperbolic graphs? We show that the minimum value of $\delta$ is $\frac{1}{2}$ using a construction of an infinite family of cactus graphs with hyperbolicity $\frac{1}{2}$. Further, we improve the lower bound of the expression $\lim_{\MP(G)\to \infty}$ $\sup\{\gamma_{b}(G)/\MP(G)\}$ for connected chordal graphs to $4/3$ where the previously best known lower bound was $10/9$.


\end{abstract}

 \begin{keyword}
 Cactus graph \sep Hyperbolic graph \sep Multipacking \sep Dominating broadcast.
 \end{keyword}
\end{frontmatter}

\section{Introduction}
\label{sec:introduction}


Covering and packing  are fundamental problems in graph theory and algorithms~\cite{cornuejols2001combinatorial}.  In this paper, we study two dual covering and packing problems called \emph{broadcast domination} and \emph{multipacking}. The broadcast domination problem is motivated by telecommunication networks. Imagine a network with radio towers that can transmit information within a certain radius 
 $r$ for a cost of $r$. The goal is to cover the entire network while minimizing the total cost. The multipacking problem is its natural packing counterpart and generalizes various other standard packing problems. Unlike many standard packing and covering problems, these two problems involve arbitrary distances in graphs, which makes them challenging. The goal of this paper is to study the relation between these two parameters in the class of cactus graphs.

For a graph $ G = (V, E) $ with  vertex set $ V $,  edge set $ E $ and the diameter $diam(G)$, a function
	$ f : V \rightarrow \{0, 1, 2, . . . , diam(G)\} $ is called a \textit{broadcast} on $ G $. Suppose $G$ is a graph with a broadcast $f$. Let $d(u,v)$ be the length of a shortest path joining the vertices $u$ and $v$ in $G$.  We say $v\in V$ is a \textit{tower} of $G$ if $f(v)>0$.	
Suppose $u, v  \in  V$ (possibly, $ u = v $) such that $ f (v) > 0 $ and
	$ d(u, v) \leq f (v) $, then we say $v$ \textit{broadcasts} (or  \textit{dominates}) $u$, and $u$ \textit{hears} the broadcast from $v$.

For each
	vertex $ u \in V  $, if  there exists a vertex $ v $ in $ G $ (possibly, $ u = v $) such that $ f (v) > 0 $ and
	$ d(u, v) \leq f (v) $, then $ f $ is called a \textit{dominating broadcast} on $ G $.
The \textit{cost} of the broadcast $f$ is the quantity $ \sigma(f)  $, which is the sum
	of the weights of the broadcasts over all vertices in $ G $. So, $\sigma(f)=  \sum_{v\in V}f(v)$. The minimum cost of a dominating broadcast in G (taken over all dominating broadcasts)  is the \textit{broadcast domination number} of G, denoted by $ \gamma_{b}(G) $.  So, $ \gamma_{b}(G) = \displaystyle\min_{f\in D(G)} \sigma(f)= \displaystyle\min_{f\in D(G)} \sum_{v\in V}f(v)$, where $D(G)$ is the set of all dominating broadcasts on $G$. 
	
	Suppose $f$ is a dominating broadcast with $f(v)\in \{0,1\}$ for each $ v\in V(G)$, then $\{v\in V(G):f(v)=1\}$ is a \textit{dominating set} on $G$. The minimum cardinality of a dominating set is the \textit{domination number} which is denoted by $ \gamma(G) $.

An \textit{optimal broadcast} or \textit{optimal dominating broadcast} on a graph $G$ is a dominating broadcast with a cost equal to $ \gamma_{b}(G) $.	
A dominating broadcast is \textit{efficient} if no vertex
hears a broadcast from two different vertices. Therefore, no tower can hear a broadcast from another tower in an efficient broadcast. There is a theorem that says,	for every graph there is an optimal efficient dominating broadcast \cite{dunbar2006broadcasts}. 
	Define a ball of radius $r$ around $v$ by $N_r[v]=\{u\in V(G):d(v,u)\leq r\}$.  Suppose $V(G)=\{v_1,v_2,v_3,\dots,v_n\}$. Let $c$ and $x$ be the vectors indexed by $(i,k)$ where $v_i\in V(G)$ and $1\leq k\leq diam(G)$, with the  entries $c_{i,k}=k$ and $x_{i,k}=1$ when $f(v_i)=k$ and $x_{i,k}=0$ when $f(v_i)\neq k$. Let $A=[a_{j,(i,k)}]$ be a matrix with the entries 
\begin{center}	$a_{j,(i,k)}=
    \begin{cases}
        1 & \text{if }  v_j\in N_k[v_i]\\
        0 & \text{otherwise. }
    \end{cases} $
 \end{center}

	Hence, the broadcast domination number can be expressed as an integer linear program:  
 \begin{center}
	    $\gamma_b(G)=\min \{c\cdot x :  Ax\geq \mathbf{1}, x_{i,k}\in \{0,1\}\}.	$
     \end{center}
	
  The \textit{maximum multipacking problem} is the dual integer program of the above problem. In 2013, Brewster, Mynhardt, and Teshima \cite{brewster2013new} obtained a
generalization of 2-packings called multipackings. A \textit{k-multipacking} is a set $ M \subseteq V  $ in a
	graph $ G = (V, E) $ such that   $|N_r[v]\cap M|\leq r$ for each vertex $ v \in V(G) $ and for every integer $ 1\leq r \leq k $. The \textit{k-multipacking number} of $ G $ is the maximum cardinality of a $k$-multipacking of $ G $ and it
	is denoted by $ \MP_k(G) $. If $M$ is a $k$-multipacking, where $k=diam(G)$, the diameter of $G$, then $M$ is called a \textit{multipacking} of $G$, and the $k$-multipacking number of $G$ is called \textit{multipacking number} of $G$, denoted by $\MP(G)$.  A \textit{maximum multipacking} is a multipacking $M$  of a graph $ G  $ such that	$|M|=\MP(G)$. If $M$ is a multipacking, we define   a vector $y$ with the entries $y_j=1$ when $v_j\in M$ and $y_j=0$ when $v_j\notin M$.  So,  \begin{center}
	    $\MP(G)=\max \{y\cdot\mathbf{1} :  yA\leq c, y_{j}\in \{0,1\}\}.$ 
\end{center}


A $k$-packing is a set of vertices in $G$, such that the shortest path between each pair of the vertices from the set  has at least $k +1$ edges. When $k =1$, the $1$-packing set problem is called the independent set problem.
 The usual $2$-packing is a $1$-multipacking of a graph. 

\medskip
\noindent\textbf{Brief Survey: }  Packing is a well-studied research topic in graph theory.    In 1985, Hochbaum, and Shmoys~\cite{hochbaum1985best} proved that finding a maximum $k$-packing set in an arbitrary graph is an NP-hard problem, for any $k$.  Very few graph classes have polynomial time algorithms to solve this kind of problems (e.g., rings and trees) \cite{mjelde2004k}.  In 2018, Flores-Lamas, Fern{\'a}ndez-Zepeda, and Trejo-S{\'a}nchez~\cite{flores2018algorithm} provided a polynomial time algorithm to find a maximum 2-packing set in a cactus.  
The application areas of these problems include  information retrieval, classification theory, computer vision, biomedical engineering, scheduling, experimental design, and financial markets. Butenko \cite{butenko2003maximum} discussed these applications in detail.

 Multipacking was formally introduced in Teshima’s Master’s Thesis \cite{teshima2012broadcasts} in 2012 (also see \cite{beaudou2019multipacking,cornuejols2001combinatorial,dunbar2006broadcasts,meir1975relations}). 
However, until now, there is no known polynomial-time algorithm to find a maximum multipacking of general graphs, and the problem is also not known to be NP-hard. However, there are polynomial-time algorithms for trees and more generally, strongly chordal graphs~\cite{brewster2019broadcast} to solve the multipacking problem. See \cite{das2024multipacking} for the geometric version of this problem.

Broadcast domination is a generalization of domination problems. Erwin~\cite{erwin2004dominating,erwin2001cost} introduced broadcast domination in his doctoral thesis in
2001. For general graphs, an optimal dominating broadcast can be found in polynomial-time $O(n^6)$~\cite{heggernes2006optimal}. The same problem can be solved in linear time for trees~\cite{brewster2019broadcast}. See~\cite{foucaud2021complexity} for other references concerning algorithmic results on these problems.

It is known that $\MP(G)\leq \gamma_b(G)$~\cite{brewster2013new}. In 2019, Beaudou,  Brewster, and Foucaud~\cite{beaudou2019broadcast} proved that $\gamma_b(G)\leq 2\MP(G)+3$ and they conjectured that $\gamma_b(G)\leq 2\MP(G)$ for every graph. Moreover, they provided an approximation algorithm that constructs a multipacking of a general graph $G$ of size at least $\frac{\MP(G)-3}{2}$. Hartnell and Mynhardt~\cite{hartnell2014difference} constructed a family of connected graphs such that the difference $\gamma_b(G)-\MP(G)$ can be arbitrarily large and in fact, for which the ratio $\gamma_b(G)/ \MP(G)=4/3$. Therefore, for general connected graphs, 
\begin{center}
    $\dfrac{4}{3}\leq\displaystyle\lim_{\MP(G)\to \infty}\sup\Bigg\{\dfrac{\gamma_{b}(G)}{\MP(G)}\Bigg\}\leq 2.$
\end{center} 
A natural question comes to mind: What is the optimal bound on this ratio for other graph classes? It is known that, for any connected chordal graph $G$, $\gamma_{b}(G)\leq \big\lceil{\frac{3}{2} \MP(G)\big\rceil}$~\cite{das2023relation}. It is also known  that $\gamma_b(G)-\MP(G)$ can be arbitrarily large for connected chordal graphs~\cite{das2023relation}.

\medskip
\noindent\textbf{Our Contribution: } A \textit{cactus} is a connected graph in which any two  cycles have at most one vertex in common. Equivalently, it is a connected graph in which every edge belongs to at most one  cycle. Cactus is a superclass of tree and a subclass of outerplanar graph. In this paper, we study the multipacking problem on the cactus graph. We  establish a relation between multipacking and dominating broadcast on the same graph class. 
We start by bounding the multipacking number of a cactus:

\begin{restatable}{theorem}{multipackingbroadcastrelation}
\label{thm:multipacking_broadcast_relation}
    Let $G$ be a cactus, then $\gamma_b(G)\leq \frac{3}{2}\MP(G)+\frac{11}{2}$.
\end{restatable} 

Note that, for $\delta$-hyperbolic graphs (defined in this section later),  $\gamma_{b}(G)\leq \big\lfloor{\frac{3}{2} \MP(G)+2\delta\big\rfloor} $~\cite{das2023relationarxiv}. Chordal graphs\footnote{A \textit{chordal graph} is an undirected simple graph in which all cycles of four or more vertices have a chord, which is an edge that is not part of the cycle but connects two vertices of the cycle.} are $1$-hyperbolic \cite{brinkmann2001hyperbolicity}. Earlier we mentioned that  for any connected chordal graph $G$, $\gamma_{b}(G)\leq \big\lceil{\frac{3}{2} \MP(G)\big\rceil}$~\cite{das2023relation}. The cactus graphs are far from  being chordal or  hyperbolic since cactus graphs can have unbounded hyperbolicity and that shows the importance of Theorem \ref{thm:multipacking_broadcast_relation}.  The proof of Theorem \ref{thm:multipacking_broadcast_relation} is based on finding some paths and a cycle (if needed) in the graph whose total size is almost double the radius. The set of every third vertex on these paths and the cycle yields a multipacking under some conditions. 



The hardness problem of multipacking  
has been repeatedly addressed by numerous authors, yet it has persisted as an unsolved challenge for the past decade. However, polynomial-time algorithms are known for trees and more generally, strongly chordal graphs~\cite{brewster2019broadcast}. Even for trees, the algorithm for finding a maximum multipacking   is very non-trivial. The complexity of the multipacking problem for the  planar graphs is also unknown till now. Moreover, there is no approximation algorithm for the planar graphs. Cactus is a subclass of the planar graph class.  We have already mentioned that, there is a polynomial time algorithm to find a maximum 2-packing set in a cactus \cite{flores2018algorithm}. A multipacking is a 2-packing, but the reverse is not true. In this paper, we provide an approximation algorithm to find multipacking on cactus.  Our proof technique is based on the basic structure of the cactus graphs, and the technique is completely different from the existing polynomial time solution for the $2$-packing problem on the cactus.

\begin{restatable}{theorem}{MultipackingAlgorithm}\label{thm:multipacking_algorithm}
 If $G$ is a cactus graph, there is an $O(n)$-time algorithm to construct a multipacking of $G$ of size at least $ \frac{2}{3}\MP(G)-\frac{11}{3} $  where $n=|V(G)|$.
\end{restatable}

\begin{figure}[ht]
    \centering
    \includegraphics[height=3.5cm]{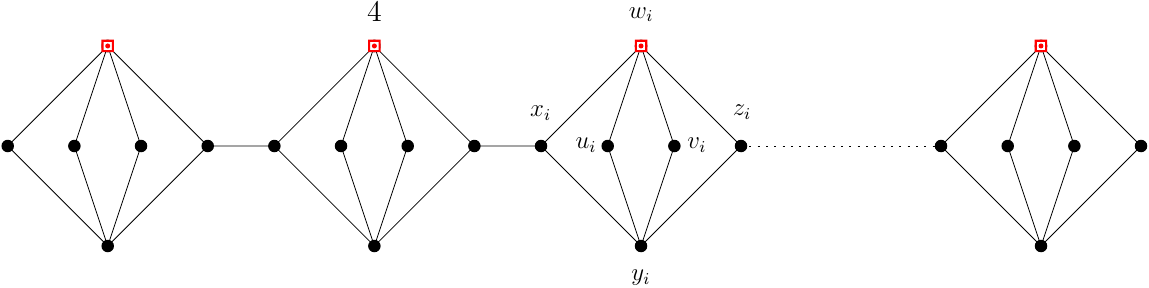}
    \caption{The $H_k$ graph with $\gamma_b(H_k)=4k$ and $\MP(H_k)=3k$. The set $\{w_i:1\leq i\leq 3k\}$ is a maximum multipacking of $H_k$. This family of graphs was constructed by Hartnell and Mynhardt~\cite{hartnell2014difference}.} 
    \label{fig:Hk}
\end{figure}

Hartnell and Mynhardt~\cite{hartnell2014difference} constructed a family of connected graphs $H_k$ such that $\MP(H_k)=3k$ and $\gamma_b(H_k)=4k$ (Fig. \ref{fig:Hk}). There is no such construction for trees and more generally, strongly chordal graphs, since multipacking number and broadcast domination number are the same for these graph classes \cite{brewster2019broadcast}. We provide a simpler family of connected graphs $G_k$ such that $\MP(G_k)=3k$ and $\gamma_b(G_k)=4k$ (Fig. \ref{fig:pentagon}). Not only that, the family of graphs $G_k$ covers many graph classes, including   cactus (a subclass of outerplanar graphs), AT-free\footnote{An independent set of three vertices such that each pair is joined by a path that avoids the neighborhood of the third is called an \textit{asteroidal triple}. A graph is asteroidal triple-free or AT-free if it contains
no asteroidal triples.} (a subclass of $C_n$-free graphs, for $n\geq 6$) graphs, etc. We state that in the following theorem:

\begin{restatable}{theorem}{multipackingbroadcastgape}\label{thm:3k4k}
 For each positive integer $k$, there is a cactus graph (and AT-free graph) $G_k$ such that $\MP(G_k)=3k$ and $\gamma_b(G_k)=4k$.
\end{restatable}


Theorem \ref{thm:3k4k} directly establishes the following corollary.

\begin{restatable}{corollary}{gammabDIFFGmpG}\label{cor:gammabG-mpG} The difference $ \gamma_{b}(G) -  \MP(G) $ can be arbitrarily large for cactus graphs (and for AT-free graphs).
\end{restatable}

We also make a connection with the \emph{fractional} versions of the two concepts dominating broadcast and multipacking, as introduced in~\cite{brewster2013broadcast}.

As mentioned earlier, for general connected graphs, the range of the expression $\lim_{\MP(G)\to \infty}$ $\sup\{\gamma_{b}(G)/\MP(G)\}$ is $[4/3,2]$.
It might be very difficult to find the exact value of this expression. For trees the value is $1$ since $\gamma_b(G)=\MP(G)$ holds for trees, and more generally for strongly chordal graphs\footnote{A graph is \textit{strongly chordal} if it is chordal and every cycle of even length ($\geq 6$) has an odd chord i.e., an edge that connects two vertices that are an odd distance ($>1$) apart from each other in the cycle.}~\cite{brewster2019broadcast}. We tried to find a more general graph class that has the range of the expression $\lim_{\MP(G)\to \infty}\sup\{\gamma_{b}(G)/\MP(G)\}$ tighter than the existing bound $[4/3,2]$. We found that the cactus graph has a tighter bound $[4/3,3/2]$ for the expression. This indicates that the bound for the general graph can be improved further. Theorem \ref{thm:multipacking_broadcast_relation} and Theorem \ref{thm:3k4k} yield the following:

\begin{restatable}{corollary}{gammabGBYmpG} \label{cor:gammabG/mpG} For cactus graphs $G$, \text{ }
$\displaystyle\frac{4}{3}\leq\lim_{\MP(G)\to \infty}\sup\Bigg\{\frac{\gamma_{b}(G)}{\MP(G)}\Bigg\}\leq \frac{3}{2}.$
\end{restatable}  

Corollary \ref{cor:gammabG/mpG} yields that, for cactus graphs, we cannot form a bound in the form $\gamma_b(G)\leq c_1\cdot \MP(G)+c_2$, for any constant $c_1<4/3$ and $c_2$.


We also answer some questions of the relation between broadcast domination number and multipacking number on the hyperbolic graph classes. Gromov~\cite{gromov1987hyperbolic} introduced the $\delta$-hyperbolicity measures to explore the geometric group theory through Cayley graphs. Let $d$ be the shortest-path metric of a graph $G$. The graph $G$ is called a \emph{$\delta$-hyperbolic graph} if for any four
vertices $u, v, w, x \in V(G)$, the two larger of the three sums $d(u, v) + d(w, x)$, $d(u, w) + d(v, x)$, $d(u, x) +
d(v, w)$ differ by at most $2\delta$. A graph class $\mathcal{G}$ is said to be hyperbolic if there exists a constant $\delta$ such that every graph $G \in \mathcal{G}$ is $\delta$-hyperbolic. It is known that a graph is $0$-hyperbolic iff it is a block graph\footnote{A graph is a \textit{block graph} if every block (maximal 2-connected component) is a clique.}~\cite{howorka1979metric}. Trees are $0$-hyperbolic graphs, since trees are block graphs. Moreover, it is known that all strongly chordal graphs are $\frac{1}{2}$-hyperbolic~\cite{wu2011hyperbolicity}, and also $\MP(G)=\gamma_b(G)$ holds for the strongly chordal graphs~\cite{brewster2019broadcast}. Since block graphs are strongly chordal graphs, therefore $\MP(G)=\gamma_b(G)$ holds for the graphs with hyperbolicity $0$. Furthermore, we know that $ \gamma_{b}(G) -  \MP(G) $ can be arbitrarily large for connected chordal graphs which is a subclass of $1$-hyperbolic graphs.  This leads to a natural question: what is the minimum value of $\delta$, for which we can say that the difference $ \gamma_{b}(G) -  \MP(G) $ can be arbitrarily large for $\delta$-hyperbolic graphs? We answer this question. First we show that the family of graphs $G_k$ (Fig. \ref{fig:pentagon}) is $\frac{1}{2}$-hyperbolic. This yields the following theorem.

\begin{restatable}{theorem}{hypermultipackingbroadcastgape}\label{thm:1/2-Hyperbolic_graphs}
For each positive integer $k$, there is a $\frac{1}{2}$-hyperbolic graph $G_k$ such that $\MP(G_k)=3k$ and $\gamma_b(G_k)=4k$.
    
\end{restatable}

Theorem \ref{thm:1/2-Hyperbolic_graphs} directly establishes the following corollary.

\begin{restatable}{corollary}{hypergammabDIFFGmpG}\label{cor:hyper_gammabG-mpG} The difference $ \gamma_{b}(G) -  \MP(G) $ can be arbitrarily large for $\frac{1}{2}$-hyperbolic graphs.
\end{restatable}

We discussed earlier that for all $0$-hyperbolic graphs, we have $ \gamma_{b}(G) =  \MP(G) $. Therefore, Corollary \ref{cor:hyper_gammabG-mpG} yields the following.

\begin{restatable}{theorem}{mindelta}\label{thm:min_delta} For $\delta$-hyperbolic graphs, the minimum value of $\delta$ is $\frac{1}{2}$ for which, the difference $ \gamma_{b}(G) -  \MP(G) $ can be arbitrarily large.
\end{restatable}

It is known that, if $G$ is a $\delta$-hyperbolic graph, then $\gamma_{b}(G)\leq \big\lfloor{\frac{3}{2} \MP(G)+2\delta\big\rfloor} $~\cite{das2023relationarxiv}. This result, together with Theorem \ref{thm:1/2-Hyperbolic_graphs}, yields the following.

\begin{restatable}{corollary}{hypergammabGBYmpG} \label{cor:hyper_gammabG/mpG} For $\frac{1}{2}$-hyperbolic graphs $G$, \text{ }
$\displaystyle\frac{4}{3}\leq\lim_{\MP(G)\to \infty}\sup\Bigg\{\frac{\gamma_{b}(G)}{\MP(G)}\Bigg\}\leq \frac{3}{2}.$
\end{restatable}  

Previously it was known that, for connected chordal graphs, the range of the expression $\lim_{\MP(G)\to \infty}$ $\sup\{\gamma_{b}(G)/\MP(G)\}$ is $[10/9,3/2]$  \cite{das2023relation}. In this paper, we improve the lower bound of the expression to $4/3$. To prove this, we have shown the following.

\begin{restatable}{theorem}{chordalmultipackingbroadcastgape}\label{thm:chordal_graphs}
For each positive integer $k$, there is a connected chordal graph $F_k$ such that $\MP(F_k)=3k$ and $\gamma_b(F_k)=4k$.
    
\end{restatable}

\noindent\textbf{Organisation:} 
In Section \ref{sec:inequality_linking_Broadcast domination_and_Multipacking}, we prove Theorem \ref{thm:multipacking_broadcast_relation}. In Section \ref{sec:approximation_algorithm_to_find_Multipacking}, we provide a $(\frac{3}{2}+o(1))$-factor approximation algorithm for finding multipacking on the cactus graphs. In Section \ref{sec:Unboundedness_of_the_gap_between_Broadcast_domination_and_Multipacking}, we prove  that the difference $ \gamma_{b}(G) -  \MP(G) $ can be arbitrarily large for cactus graphs. In Section \ref{sec:1/2-Hyperbolic graphs},  we show that the difference $ \gamma_{b}(G) -  \MP(G) $ can be arbitrarily large for $\frac{1}{2}$-hyperbolic graphs also. In Section \ref{sec:chordal graphs}, we show that, for connected chordal graphs, the lower bound of the expression $\lim_{\MP(G)\to \infty}$ $\sup\{\gamma_{b}(G)/\MP(G)\}$ is  $4/3$. We conclude in Section \ref{sec:conclusion}.

\section{Definitions and notations}

 Let $G=(V,E)$ be a graph and $d_G(u,v)$ be the length of a shortest path joining two vertices $u$ and $v$ in  $G$, we simply write $d(u,v)$ when there is no confusion. Let $diam(G)=\max\{d(u,v):u,v\in V(G)\}$. Diameter is a path of $G$  of the length  $diam(G)$. 
 $N_r[u]=\{v\in V:d(u,v)\leq r\}$ where $u\in V$. The \textit{eccentricity} $e(w)$  of a vertex $w$ is $\min \{r:N_r[w]=V\}$. The \textit{radius} of the graph $G$ is $\min\{e(w):w\in V\}$, denoted by $\rad(G)$.  The \textit{center set} $C(G)$ of the graph $G$  is the set of all vertices of minimum eccentricity, i.e., $C(G)=\{v\in V:e(v)=\rad(G)\}$. Each vertex in the set $C(G)$ is called a \textit{center} of the graph $G$.  
 A subgraph $H$ of a graph $G$ is called an \textit{isometric subgraph} if $d_H(u, v) = d_G(u, v)$ for every pair of vertices 
$u$ and $v$ in $H$, where $d_H(u,v)$ and $d_G(u,v)$ are the distances between $u$ and $v$ in $H$ and $G$, respectively. An \textit{isometric path} is an isometric subgraph which is a path.  If $H_1$ and $H_2$ are two subgraphs of $G$, then $H_1\cup H_2$ denotes the subgraph whose vertex set is $V(H_1)\cup V(H_2)$ and edge set is $E(H_1)\cup E(H_2)$. We denote an  indicator function as  $1_{[x<y]}$  that takes the value $1$ when $x<y$, and $0$ otherwise. If $P$ is a path in $G$, then we say $V(P)$ is the vertex set of the path $P$, $E(P)$ is the edge set of the path $P$, and  $l(P)$ is the length of the path $P$, i.e., $l(P)=|E(P)|$.

\section{Proof of Theorem \ref{thm:multipacking_broadcast_relation}}\label{sec:inequality_linking_Broadcast domination_and_Multipacking}

In this section, we prove a relation between the broadcast domination number and the radius of a cactus. Using this we establish our main result that relates the  broadcast domination number and the multipacking number for cactus graphs.

We start with a lemma that is true for general graphs.


\begin{lemma} \label{lem:disjoint_radial_path} \textbf{(Disjoint radial path lemma)}
Let $G$ be a graph with radius $r$ and center $c$, where $r\geq 1$. Let $P$ be an isometric path in $G$ such that $l(P)=r$ and $c$ is one endpoint of $P$. Then there exists an isometric path $Q$ in $G$ such that $V(P)\cap V(Q)=\{c\}$, $ r-1 \leq l(Q)\leq  r$ and $c$ is one endpoint of $Q$.
\end{lemma}





The main idea of proving Lemma \ref{lem:disjoint_radial_path} is the following: if the length of all the isometric paths (except $P$) with one endpoint $c$  has length less than $r-1$, then the radius of $G$ will be less than $r$. In other case,  if all the isometric paths (except $P$) of length at least $r-1$ with one endpoint $c$     share at least one common vertex except $c$ with the path $P$, then also the radius of $G$ will be less than $r$. In both cases, we arrive at a contradiction.  The formal proof is the following.

\begin{proof}[Proof of Lemma \ref{lem:disjoint_radial_path}]

    Since $\rad(G)=r$, there exists a vertex $v_r$ such that $d(c,v_r)=r$. Let $P=(c,v_1,v_2,v_3,$ $\dots ,v_r)$ be an $r$ length path that joins $c$ and $v_r$. Therefore $P$ is an isometric path of $G$.  Let $Z_k$ be the set of all isometric paths of length $k$ whose one end vertex is $c$.  Since $d(c,v_{r-1})=r-1$, it implies $Z_{r-1}\neq \phi$. We prove this lemma using contradiction. We show that if  $(V(P)\cap V(Q))\setminus \{c\}\neq \phi$ for all $Q\in Z_{r-1}$, then $\rad(G)\leq r-1$.  
    
    Suppose  $(V(P)\cap V(Q))\setminus \{c\}\neq \phi$ for all $Q\in Z_{r-1}$.
    Let $V_k=\{u\in V(G):d(c,u)=k\}$. So, $ V(G)=\bigcup_{k=0}^{r}V_k$.  
     Let $w_r\in V_r$ and $P_1$ be a shortest path joining $c$ and $w_r$. Let $P_1=(c,w_1,w_2,w_3,\dots, w_r)$ and  $P_1'=(c,w_1,w_2,w_3,\dots, w_{r-1})$. So, $P_1'\in Z_{r-1}$. Therefore, $(V(P)\cap V(P_1'))\setminus \{c\}\neq \phi$. Let $w_t\in (V(P)\cap V(P_1'))\setminus \{c\}$ where $1\leq t \leq r-1$. Since $w_t,v_t\in V(P)\cap V_t$, so $w_t=v_t$. Now consider the path $P_1''=(v_1,v_2,\dots, v_t,w_{t+1},w_{t+2},\dots, w_r)$. Here $l(P_1'')=r-1$. Therefore $d(v_1,w_r)\leq r-1$. So, $d(v_1,w)\leq r-1$ for all $w\in V_r$. Let $u_{r-1}\in V_{r-1}$ and $P_2$ be a shortest path joining $c$ and $u_{r-1}$. Let $P_2=(c,u_1,u_2,u_3,\dots, u_{r-1})$. So, $P_2\in Z_{r-1}$. Therefore $(V(P)\cap V(P_2))\setminus \{c\}\neq \phi$. Let $u_s\in  (V(P)\cap V(P_2))\setminus \{c\}$ where $1\leq s\leq r-1$. Since $u_s,v_s\in V(P)\cap V_s$, so $u_s=v_s$. Now consider the path $P_2'=(v_1,v_2,\dots, v_s,u_{s+1},u_{s+2},\dots, u_{r-1})$. Here $l(P_2')=r-2$. Therefore $d(v_1,u_{r-1})\leq r-2$. So, $d(v_1,u)\leq r-2$ for all $u\in V_{r-1}$. Since $c$ and $v_1$ are adjacent, we can say that  $d(v_1,x)\leq r-1$ for all   $x\in  \bigcup_{k=0}^{r-2}V_k$. Therefore $d(v_1,x)\leq r-1$ for all $x\in V(G)$. This implies that $\rad(G)\leq r-1$. This is a contradiction. Therefore, there exists a path $Q\in Z_{r-1}$ such that $(V(P)\cap V(Q))\setminus \{c\}= \phi$.     
\end{proof}

\begin{figure}[ht]
    \centering
    \includegraphics[scale = 0.8]{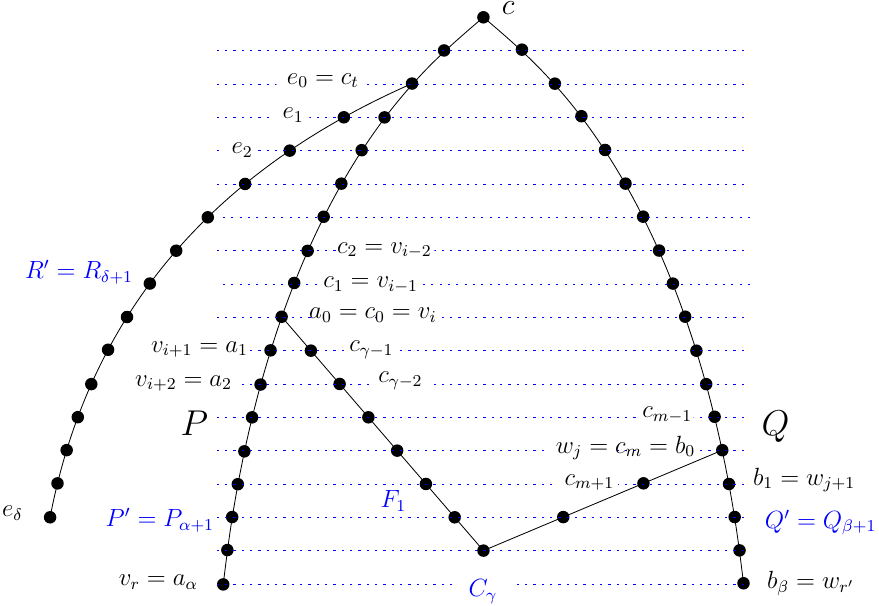}
    \caption{The subgraph $P\cup Q\cup F_1\cup R'=H_\gamma (c_0,\alpha,c_t,\delta,c_m,\beta)$}
    \label{fig:bfs3}
\end{figure}

For this section, we assume that $G$ is a cactus. 
Let $c$ be a center  and $r$ be the radius of $G$ where $r\geq 1$. We can find an isometric path $P$ of length $r$ whose one endpoint is $c$, since  $\rad(G)=r$. Let $P=(c,v_1,v_2,v_3,\dots ,v_r)$. Therefore, we can find an isometric path $Q$ in $G$ such that $V(P)\cap V(Q)=\{c\}$, $ r-1 \leq l(Q)\leq  r$ and $c$ is one endpoint of $Q$ by Lemma \ref{lem:disjoint_radial_path}. Let $Q=(c,w_1,w_2,w_3,\dots ,w_{r'})$ where $ r-1 \leq r'\leq  r$. Now we define some expressions to study the subgraph $P\cup Q$ in $G$ (See Fig. \ref{fig:bfs3}).  For $v_i\in V(P)\setminus \{c\}$ and $w_j\in V(Q)\setminus \{c\}$, we define  $ X_{P,Q}(v_i,w_j)=$\{$P_1$ :  $P_1$ is a path in $G$ that joins $v_i$ and $w_j$ such that $V(P)\cap V(P_1)=\{v_i\}$, $V(Q)\cap V(P_1)=\{w_j\}$ and $c \notin V(P_1)$\}. Let $X_{P,Q}=\{(v_i,w_j): X_{P,Q}(v_i,w_j)\neq \phi \}$. Since $G$ is a cactus, it implies every edge belongs to at most one cycle. Therefore, $|X_{P,Q}(v_i,w_j)|\leq 1$ and also $|X_{P,Q}|\leq 1$. Therefore, there is at most one path (say, $P_1$) that does not pass through $c$ and joins a vertex of $V(P)\setminus \{c\}$ with a vertex of $V(Q)\setminus \{c\}$ and $P_1$ intersects $P$ and $Q$ only at their joining points. So, the following observation is true. 

\begin{observation}
\label{obs:joining_path_atmost_1}
    Let $G$ be a cactus with $\rad(G)=r$ and center $c$. Suppose $P$ and $Q$  are two isometric paths  in $G$ such that   $V(P)\cap V(Q)=\{c\}$,  $l(P)=r$, $r-1\leq l(Q)\leq r$  and both have  one endpoint  $c$. Then 
    
    (i) $|X_{P,Q}|\leq 1$ and  $|X_{P,Q}(v_i,w_j)|\leq 1$ for all $(v_i,w_j)$.

(ii) $X_{P,Q}= \{(v_i,w_j)\}$ iff  $|X_{P,Q}(v_i,w_j)|= 1$.
    
\end{observation}

\begin{figure}[ht]
    \centering
    \includegraphics[height=6.5cm]{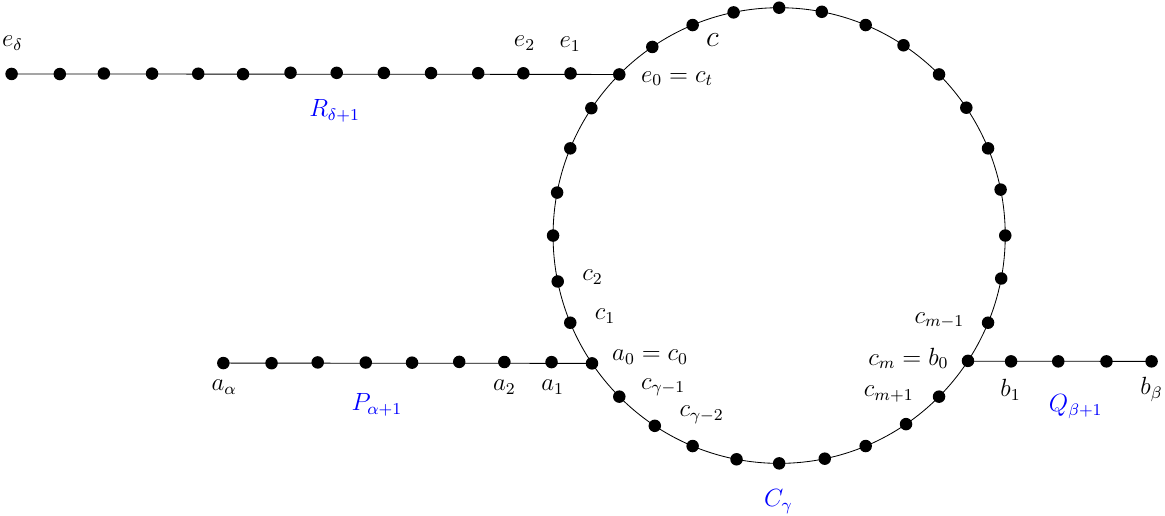}
    \caption{The subgraph $H_\gamma (c_0,\alpha,c_t,\delta,c_m,\beta)$}
    \label{fig:bfs1}
\end{figure}

\medskip
\noindent \textbf{Finding a special subgraph $\bm{H_\gamma(c_0,\alpha,c_t,\delta,c_m,\beta)}$  in cactus $\bm{G}$ : }
From Observation \ref{obs:joining_path_atmost_1}, there are two possible cases: either $|X_{P,Q}| =0$ or $|X_{P,Q}|=1$. First, we want to study the case when $|X_{P,Q}|=1$. In that case, suppose $ X_{P,Q}=\{(v_i,w_j)\}$ and $X_{P,Q}(v_i,w_j)=\{F_1\}$ (See Fig.  \ref{fig:bfs3}). We are interested in finding a multipacking set of $G$ from the subgraph $P\cup Q\cup F_1$.   Let $F_2=(v_i,v_{i-1},v_{i-2},\dots,v_1,c,w_1,w_2,\dots,$ $w_{j-1},w_j)$, $P'=(v_i,v_{i+1},\dots,v_{r})$ and $Q'=(w_j,w_{j+1},\dots,w_{r'})$. Here $F_1\cup F_2$ is a cycle and $P'$ and $Q'$ are two isometric paths that are attached to the cycle. Suppose there is another isometric path $R'$ which is disjoint with $P'$ and $Q'$ and whose one endpoint belongs to $V(F_2)$. Then  $P\cup Q\cup F_1\cup R'=F_1\cup F_2\cup P'\cup Q' \cup R'$ is a subgraph of $G$ (See Fig.\ref{fig:bfs3}).   Let $H=P\cup Q\cup F_1\cup R'$. Note that, we can always find $P$ and $Q$ in 
$G$ due to Lemma \ref{lem:disjoint_radial_path}, but  $F_1$ or $R'$ might not be there in $G$, in that case, we can assume that $V(F_1)$ or $V(R')$ is empty in $H$. In the rest of this section, our main goal is to find a multipacking of $G$ of size at least  $\frac{2}{3}\rad(G)-\frac{11}{3}$ from the subgraph $H$, so that we can prove Theorem \ref{thm:multipacking_broadcast_relation}. To give $H$ a general structure, we want to rename the vertices. Suppose  $F_1\cup F_2$ is a cycle of length $\gamma$. Let $F_1\cup F_2=C_\gamma=(c_0,c_1,c_2,\dots,c_{\gamma-2},c_{\gamma-1},c_0)$. Suppose $l(P')=\alpha$, $l(Q')=\beta$ and $l(R')=\delta$. We rename the paths $P'$, $Q'$ and $R'$ as $P_{\alpha+1}$, $Q_{\beta+1}$ and $R_{\delta+1}$ respectively. 
Let $P_{\alpha+1}=(a_0,a_{1},\dots,a_{\alpha})$,  $Q_{\beta+1}=(b_0,b_{1},\dots,b_{\beta})$ and  $R_{\delta+1}=(e_0,e_{1},\dots,e_{\delta})$   such that $c_0=a_0$,  $c_t=e_0$, $c_m=b_0$  (See Fig. \ref{fig:bfs1}). Here $P_{\alpha+1}$, $Q_{\beta+1}$ and $R_{\delta+1}$ are three isometric paths in $G$.  According to the structure, we have $V(P_{\alpha+1})\cap V(Q_{\beta+1})=\phi$,  $V(Q_{\beta+1}) \cap V(R_{\delta+1}) =\phi$,
$V(R_{\delta+1})\cap V(P_{\alpha+1})=\phi$,
$V(C_\gamma)\cap V(P_{\alpha+1}) =\{c_0\}$, $V(C_\gamma)\cap V(R_{\delta+1})=\{c_t\}$, $V(C_\gamma)\cap V(Q_{\beta+1})=\{c_m\}$. Now we can write $H$ as a variable of $\alpha, \beta, \gamma $ and $\delta$. Let $H=H_\gamma(c_0,\alpha,c_t,\delta,c_m,\beta)$.

In the rest of the section, we study how to find a multipacking of $G$ from the subgraph $H_\gamma(c_0,\alpha,c_t,$ $\delta,c_m,\beta)$. Using this, we prove Lemma \ref{lem:multipacking_radius_relation} that yields Theorem \ref{thm:multipacking_broadcast_relation} which is our main result in this section.

\begin{lemma} [\cite{beaudou2019broadcast}]  \label{lem:isometric_path}
    Let $G$ be a graph, $k$ be a positive integer and $P=(v_0,v_1, \dots ,$ $ v_{k-1})$ be an isometric path in $G$ with $k$ vertices. Let $M = \{v_i:0\leq i\leq k, i\equiv 0 \text{ (mod $3$)} \}$ be the set of every third vertex
on this path. Then $M$ is a multipacking in $G$ of size $\big\lceil\frac{k}{3}\big\rceil$.
\end{lemma}

Note that, if $H_\gamma(c_0,\alpha,c_t,\delta,c_m,\beta)$ is a subgraph of $G$, then $P_{\alpha+1}$, $Q_{\beta+1}$ and $R_{\delta+1}$ are three isometric paths in $G$ by the definition of $H_\gamma(c_0,\alpha,c_t,\delta,c_m,\beta)$. Moreover, since $G$ is a cactus, two cycles cannot share an edge. Therefore, the following observation is true.

\begin{observation} 
\label{obs:isometric_graph}
    If $G$ is a cactus and $H_\gamma(c_0,\alpha,c_t,\delta,c_m,\beta)$ is a subgraph of $G$ such that $\gamma\geq 3 $ and $c_0,c_t,c_m$ are distinct vertices of $C_\gamma$, then $H_\gamma(c_0,\alpha,c_t,\delta,c_m,\beta)$ is an isometric subgraph of $G$.
\end{observation}

\begin{observation} 
\label{obs:isometric_graph_path_F_1}
    Let $G$ be a cactus and $H_\gamma(c_0,\alpha,c_t,\delta,c_m,\beta)$ be a subgraph of $G$ such that $\gamma\geq 3 $ and $c_0,c_t,c_m$ are distinct vertices of $C_\gamma$. Let $F_1$ and $F_2$ be two paths such that  $F_1=(c_m,c_{m+1},\dots,$ $c_{\gamma-1},c_0)$ and $F_2=(c_0,c_1,\dots,c_m)$. Then
    
 \noindent  (i)  If $l(F_1)> l(F_2)$, then $P_{\alpha+1}\cup F_2\cup Q_{\beta +1}$ is an isometric path of $G$.

 \noindent (ii)  If $l(F_1)<l(F_2)$, then $P_{\alpha+1}\cup F_1\cup Q_{\beta +1}$ is an isometric path of $G$. 

 \noindent (iii) If $l(F_1)=l(F_2)$, then 
 both of $P_{\alpha+1}\cup F_1\cup Q_{\beta +1}$ and $P_{\alpha+1}\cup F_2\cup Q_{\beta +1}$ are isometric paths of $G$.

\end{observation}

The following lemma tells why it is sufficient to find multipacking in $H_\gamma(c_0,\alpha,$ $c_t,\delta,c_m,\beta)$ to provide a multipacking in $G$.

\begin{lemma}   
\label{lem:multipacking_subgraph}
    Let $G$ be a cactus and $H_\gamma(c_0,\alpha,c_t,\delta,c_m,\beta)$ be a subgraph of $G$. If   $M$ is a multipacking of $H_\gamma(c_0,\alpha,c_t,\delta,c_m,\beta)$, then $M$ is a  
 multipacking of $G$.
\end{lemma}

\begin{proof} 
Let $H=H_\gamma(c_0,\alpha,c_t,\delta,c_m,\beta)$. Since $M$ is multipacking of $H$, therefore $|N_r[z]\cap M|\leq r$ for all $z\in V(H)$ and $r\geq 1$. Let $z\in V(G)\setminus V(H)$ and    $v\in V(H)$.  Define $ P_H(z,v)=$\{$P$ :  $P$ is a path in $G$ that joins $z$ and $v$ such that $V(P)\cap V(H)=\{v\}$\}. 
Note that, if $v_1,v_2\in V(H)$ and $v_1\neq v_2$, then $P_H(z,v_1)\cap P_H(z,v_2)=\phi$. Since $G$ is connected, therefore $|\{v\in V(H):P_H(z,v)\neq \phi\}|\geq 1$.

\medskip
\noindent
\textbf{Claim \ref{lem:multipacking_subgraph}.1. }  If $z\in V(G)\setminus V(H)$, then $|\{v\in V(H):P_H(z,v)\neq \phi\}|\leq 2$. 

\begin{claimproof}Suppose $|\{v\in V(H):P_H(z,v)\neq \phi\}|\geq 3$. Let $v_1,v_2,v_3\in \{v\in V(H):P_H(z,v)\neq \phi\}$ where $v_1,v_2,v_3$ are distinct. So, there are $3$ distinct paths $P_1,P_2,P_3$ such that $P_i\in P_H(z,v_i)$ for $i=1,2,3$. Then there are two cycles formed by the paths $P_1,P_2,P_3$ that have at least one common edge, which is a contradiction, since $G$ is a cactus.     
\end{claimproof}

\vspace{0.2cm}
\noindent
\textbf{Claim \ref{lem:multipacking_subgraph}.2. }    If $z\in V(G)\setminus V(H)$ and  $|\{v\in V(H):P_H(z,v)\neq \phi\}|=1 $, then  $|N_r[z]\cap M|\leq r$ for all $r\geq 1$.

\begin{claimproof} Let $ \{v\in V(H):P_H(z,v)\neq \phi\}=\{v_1\}$. Therefore,  if $v$ is any vertex in $V(H)$, then  any path joining $z$ and $v$ passes through $v_1$. Let $d(z,v_1)=k$ for some $k\geq 1$. We have $|N_{r}[v_1]\cap M|\leq r$ for all $r\geq 1$. If $1\leq r<k$, then  $|N_r[z]\cap M|=0<r$. If $r\geq k$, then $|N_r[z]\cap M|= |N_{r-k}[v_1]\cap M|\leq r-k\leq r$.   
\end{claimproof}

\vspace{0.2cm}
\noindent
\textbf{Claim \ref{lem:multipacking_subgraph}.3. }    If $z\in V(G)\setminus V(H)$ and $|\{v\in V(H):P_H(z,v)\neq \phi\}|=2 $, then  $|N_r[z]\cap M|\leq r$ for all $r\geq 1$.

\begin{claimproof} Let $\{v\in V(H):P_H(z,v)\neq \phi\}=\{v_1,v_2\}$. Therefore,  if $w$ is any vertex in $V(H)$, then  any path joining $z$ and $w$ passes through $v_1$ or $v_2$. Note that, both of  $v_1,v_2$ belongs to either $P_{\alpha+1}$, $Q_{\beta+1}$ or $R_{\delta+1}$, otherwise $G$ cannot be a cactus. W.l.o.g. assume that $v_1,v_2\in P_{\alpha+1}$.  Let $d(z,v_1)=k_1$ and $d(z,v_2)=k_2$ for some $k_1,k_2\geq 1$. Since $P_{\alpha+1}$ is an isometric path in $G$, therefore $d(v_1,v_2)\leq d(z,v_1)+d(z,v_2)=k_1 + k_2$. Let $r$ be a positive integer  and  $S=V(H)$. If $N_{r-k_1}[v_1]\cap N_{r-k_2}[v_2]\cap S=\phi$,  then  $(r-k_1)+(r-k_2)\leq d(v_1,v_2)\leq k_1+k_2$ $\implies$ $r\leq k_1+k_2$. Therefore,   $|N_r[z]\cap M|=|N_{r-k_1}[v_1]\cap M|+|N_{r-k_2}[v_2]\cap M|\leq r-k_1+r-k_2=2r-(k_1+k_2)\leq r$. Suppose  $N_{r-k_1}[v_1]\cap N_{r-k_2}[v_2]\cap S\neq \phi$. Let $v_1=a_i$, $v_2=a_j$, $h=\big\lfloor\frac{i+j}{2}\big\rfloor$, $v=a_h$ and $s=\big\lfloor\frac{(r-k_1)+(r-k_2)+ d(v_1,v_2)+1}{2}\big\rfloor$.  So, $N_r[z]\cap S=(N_{r-k_1}[v_1]\cup N_{r-k_2}[v_2])\cap S\subseteq N_s[v]\cap S$ $\implies$ $N_r[z]\cap M\subseteq N_s[v]\cap M$ $\implies$ $|N_r[z]\cap M|\leq |N_s[v]\cap M|\leq s= \big\lfloor\frac{(r-k_1)+(r-k_2)+ d(v_1,v_2)+1}{2}\big\rfloor\leq \big\lfloor\frac{(r-k_1)+(r-k_2)+ k_1+k_2+1}{2}\big\rfloor\leq \big\lfloor\frac{2r+1}{2}\big\rfloor= \big\lfloor r+\frac{1}{2}\big\rfloor= r$.   
\end{claimproof}

From the above results, we can say that   $|N_{r}[z]\cap M|\leq r$ for all $z\in V(G)$ and $r\geq 1$. Therefore, $M$ is a   multipacking of $G$. 
\end{proof}

Now our goal is to find multipacking of $H_\gamma(c_0,\alpha,c_t,\delta,c_m,\beta)$. Whatever multipacking we find for the subgraph $H_\gamma(c_0,\alpha,c_t,\delta,c_m,\beta)$, that will be a multipacking for $G$ by Lemma \ref{lem:multipacking_subgraph}. There could be several ways to choose a multipacking set from the subgraph $H_\gamma(c_0,\alpha,c_t,\delta,c_m,\beta)$. 
But in order to prove the Lemma \ref{lem:multipacking_radius_relation} which is the main lemma to prove Theorem \ref{thm:multipacking_broadcast_relation}, we have to ensure that the size of the multipacking is at least $\frac{2}{3}\rad(G)-\frac{11}{3}$.

We discuss three choices to find  a  multipacking in $H_\gamma(c_0,\alpha,c_t,\delta,c_m,\beta)$ in the following subsections.

From now,  we can write $H$ in place of $H_\gamma(c_0,\alpha,c_t,\delta,c_m,\beta)$ for the simplicity.

\begin{figure}[ht]
    \centering
    \includegraphics[  height=6.5cm]{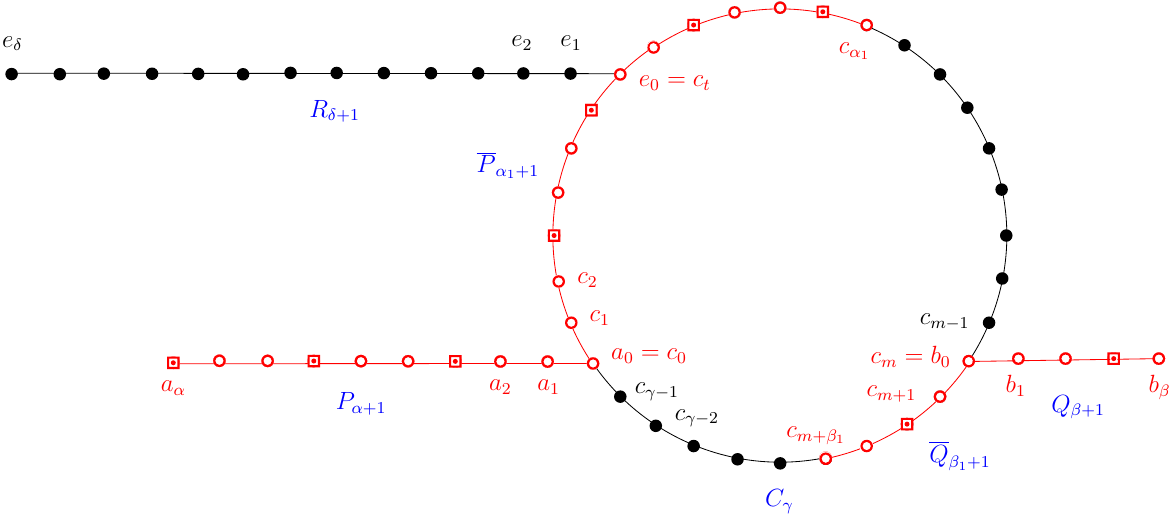}
 \caption{ The circles and squares  represent the  subgraph $P_{\alpha+1}\cup \overline{P}_{\alpha_1+1}\cup Q_{\beta+1}\cup \overline{Q}_{\beta_1+1}$ and the squares represent the set $M_\gamma(c_0,\alpha,\alpha_{1},c_m,\beta,\beta_1)$ in this figure.}
    \label{fig:m_2}
\end{figure}

\medskip
\noindent\textbf{Finding a multipacking of  $\bm{H_\gamma(c_0,\alpha,c_t,\delta,c_m,\beta)} $ 
 according to Choice-1 :}\label{subec:multipacking_technique_1}
\smallskip

Here we find two paths of  $H$ so that the set of every third vertex (with some exceptions) 
on those paths provide a multipacking of $H$. Let $\overline{P}_{\alpha_1+1}=(c_0,c_1,\dots,c_{\alpha_1})$
and $\overline{Q}_{\beta_1+1}=(c_m,c_{m+1},\dots,c_{m+\beta_1})$ where $0\leq \alpha_1\leq m-1$ and $0\leq \beta_1\leq (\gamma-1)-m$. Now we choose every third vertex (with some exceptions) from the paths  $P_{\alpha+1}\cup \overline{P}_{\alpha_1+1}$ and $Q_{\beta+1}\cup \overline{Q}_{\beta_1+1}$ to construct a set, called $M_\gamma(c_0,\alpha,\alpha_1,c_m,\beta,\beta_1)$ (See Fig. \ref{fig:m_2}). Formally we can define,  $M_\gamma(c_0,\alpha,\alpha_1,c_m,\beta,\beta_1)=\{a_i:0\leq i\leq \alpha,i\equiv 0 \text{ (mod $3$)}\}\cup \{c_i:0\leq i\leq \alpha_1, i\equiv 0 \text{ (mod $3$)}\}\cup \{b_i:0\leq i\leq \beta, i\equiv 0 \text{ (mod $3$)}\}\cup \{c_i:m\leq i\leq m+\beta_1, i\equiv m \text{ (mod $3$)}\}\setminus\{c_0,c_m\}$.  This set will be a multipacking of $G$ under the conditions stated in the following lemma.

\begin{lemma} 
\label{lem:multipacking_subgraph1}
Let $G$ be a cactus and $H_\gamma(c_0,\alpha,c_t,\delta,c_m,\beta)$  be a subgraph of $G$. Let  $\alpha_1,\alpha_2,\beta_1,\beta_2$ be non negative integers such that $\alpha_2=(m-1)-\alpha_1 $, $\beta_2=(\gamma-1)-(m+\beta_1) $. If $\alpha_1\leq 3\beta_2+\alpha_2+\beta_1$ and $\beta_1\leq 3\alpha_2+\beta_2+\alpha_1$, then 
 $M_\gamma(c_0,\alpha,\alpha_1,c_m,\beta,\beta_1)$  is a multipacking of $G$ of size at least
 $ \big\lfloor\frac{\alpha+\alpha_1+1}{3}\big\rfloor+\big\lfloor\frac{\beta+\beta_1+1}{3}\big\rfloor -2$.

\end{lemma}

\begin{proof}

Let $H=H_\gamma(c_0,\alpha,c_t,\delta,c_m,\beta)$,  $M=M_\gamma(c_0,\alpha,\alpha_1,c_m,\beta,\beta_1)$.

Let $v=c_{x_1}$ for some $x_1$ where $0\leq x_1\leq m-1$. Note that, $m-1=\alpha_1+\alpha_2$ and $\gamma=\alpha_1+\alpha_2+\beta_1+\beta_2+2$  or $\big\lfloor\frac{\gamma}{2}\big\rfloor=\big\lfloor\frac{\beta_{1}+\beta_{2}+\alpha_{1}+\alpha_2}{2}\big\rfloor+1$.

Recall that, we denote an  indicator function as  $1_{[x<y]}$  that takes the value $1$ when $x<y$, and $0$ otherwise.

\vspace{0.22cm}

 \noindent \textit{\textbf{Case 1: }} $1\leq r\leq \max\{\alpha_{1}+\alpha_2-x_1,\beta_{2}+x_1\}$. 

 $|N_r(v)\cap M|\leq \frac{1}{3} [\{2r-2(\alpha_{1}+\alpha_2-x_1)-1+(\alpha_1-x_1)+r+1\}\times 1_{[\alpha_{1}+\alpha_2-x_1\leq \beta_{2}+x_1]}+\{r-x_1-\beta_{2}+2r+1\}\times 1_{[\alpha_{1}+\alpha_2-x_1> \beta_{2}+x_1]}]\leq r $. Therefore $|N_r(v)\cap M|\leq r$.

\vspace{0.22cm}

 \noindent \textit{\textbf{Case 2: }} $\max\{\alpha_{1}+\alpha_2-x_1,\beta_{2}+x_1\}<r\leq \big\lfloor\frac{\beta_{1}+\beta_{2}+\alpha_{1}+\alpha_2}{2}\big\rfloor$. 

 $|N_r(v)\cap M|\leq \frac{1}{3} [ r+1+(\alpha_1-x_1)+2\{r-(\alpha_{1}+\alpha_2-x_1)\}-1+r-(x_1+\beta_{2})]=\frac{1}{3} [4r-\alpha_{1}-\beta_{2}-2\alpha_2]$.
We know that $\beta_{1}\leq \alpha_{1}+\beta_{2}+3\alpha_2$ $\implies$ $\beta_{1}+\beta_{2}+\alpha_{1}+\alpha_2\leq 2\alpha_{1}+2\beta_{2}+4\alpha_2$ $\implies$ $\frac{\beta_{1}+\beta_{2}+\alpha_{1}+\alpha_2}{2}\leq \alpha_{1}+\beta_{2}+2\alpha_2$. Since $r\leq \big\lfloor\frac{\beta_{1}+\beta_{2}+\alpha_{1}+\alpha_2}{2}\big\rfloor$, therefore $r\leq \alpha_{1}+\beta_{2}+2\alpha_2$  $\implies$ $4r-\alpha_{1}-\beta_{2}-2\alpha_2\leq 3r$ $\implies$ $|N_r(v)\cap M|\leq r$.

\vspace{0.22cm}

 \noindent \textit{\textbf{Case 3: }} $\big\lfloor\frac{\beta_{1}+\beta_{2}+\alpha_{1}+\alpha_2}{2}\big\rfloor< r$. 
 
 $|N_r(v)\cap M|\leq \frac{1}{3} [\beta_{1}+1+\alpha_{1}+1+(r-x_1)+r-(\alpha_{1}+\alpha_2-x_1+1)]=\frac{1}{3} [2r+\beta_{1}-\alpha_2+1]$. We know that $\beta_{1}\leq \alpha_{1}+\beta_{2}+3\alpha_2$ $\implies$ $2\beta_{1}-2\alpha_2\leq \beta_{1}+\beta_{2}+\alpha_{1}+\alpha_2$ $\implies$ $\beta_{1}-\alpha_2\leq \frac{\beta_{1}+\beta_{2}+\alpha_{1}+\alpha_2}{2}$  $\implies$ $\beta_{1}-\alpha_2+1\leq \frac{\beta_{1}+\beta_{2}+\alpha_{1}+\alpha_2}{2}+1$ $\implies$ $\beta_{1}-\alpha_2+1\leq \big\lfloor\frac{\beta_{1}+\beta_{2}+\alpha_{1}+\alpha_2}{2}\big\rfloor+1\leq r$ 
 $\implies$ $\beta_{1}-\alpha_2 +1\leq r$ $\implies$ $|N_r(v)\cap M|\leq \frac{1}{3} [2r+\beta_{1}-\alpha_2+1]\leq r$.

\vspace{0.22cm}


 Similarly, using the relation $\alpha_1\leq 3\beta_2+\alpha_2+\beta_1$, we can show that, when $v=c_{x_2}$ for some $x_2$ where $m\leq x_2\leq \gamma-1$, we can show that  $v\in \{c_i:m\leq i\leq \gamma-1\}$, then $|N_r(v)\cap M|\leq r$ for all  $r\geq 1$. Therefore, $|N_r(v)\cap M|\leq r$ for all $v\in V(C_\gamma)$ and $r\geq 1$.

 Suppose $v\in V(P_{\alpha+1})$, then any path that joins $v$ with a vertex in $V(C_\gamma)\cup V(Q_{\beta+1})\cup V(R_{\delta+1})$ passes through $a_0$, otherwise $G$ cannot be a cactus. By Observation \ref{obs:isometric_graph} we know that $H_\gamma(c_0,\alpha,c_t,\delta,c_m,\beta)$ is an isometric subgraph of $G$.  Therefore $|N_r(v)\cap M|\leq r$ for all  $r\geq 1$. Similarly we can show that, if $v$ is in  $V(Q_{\beta+1})$ or $V(R_{\delta+1})$, then $|N_r(v)\cap M|\leq r$ for all  $r\geq 1$. Therefore $M$ is a multipacking of $H$. So, $M$ is a multipacking of $G$ by Lemma \ref{lem:multipacking_subgraph} and $|M|\geq  \big\lfloor\frac{\alpha+\alpha_1+1}{3}\big\rfloor+\big\lfloor\frac{\beta+\beta_1+1}{3}\big\rfloor -2$.   
\end{proof}

Substitute $\alpha_2=(m-1)-\alpha_1 $ and $\beta_2=(\gamma-1)-(m+\beta_1) $ in Lemma \ref{lem:multipacking_subgraph1}. This yields the following.

\begin{lemma} 
\label{lem:gamma/2}
    Let $G$ be a cactus and $H_\gamma(c_0,\alpha,c_t,\delta,c_m,\beta)$  be a subgraph of $G$. Let  $\alpha_1$ and $\beta_1$ be non negative integers such that $\alpha_1\leq m-1$ and $\beta_1\leq (\gamma-1)-m$. If $\alpha_1\leq \big\lfloor\frac{\gamma}{2}\big\rfloor-1$ and $\beta_1\leq \big\lfloor\frac{\gamma}{2}\big\rfloor-1$, then 
 $M_\gamma(c_0,\alpha,\alpha_1,c_m,\beta,\beta_1)$  is a multipacking of $G$ of size at least
 $ \big\lfloor\frac{\alpha+\alpha_1+1}{3}\big\rfloor+\big\lfloor\frac{\beta+\beta_1+1}{3}\big\rfloor -2$.
 
\end{lemma}



\begin{figure}[ht]
    \centering
    \includegraphics[  height=6.5cm]{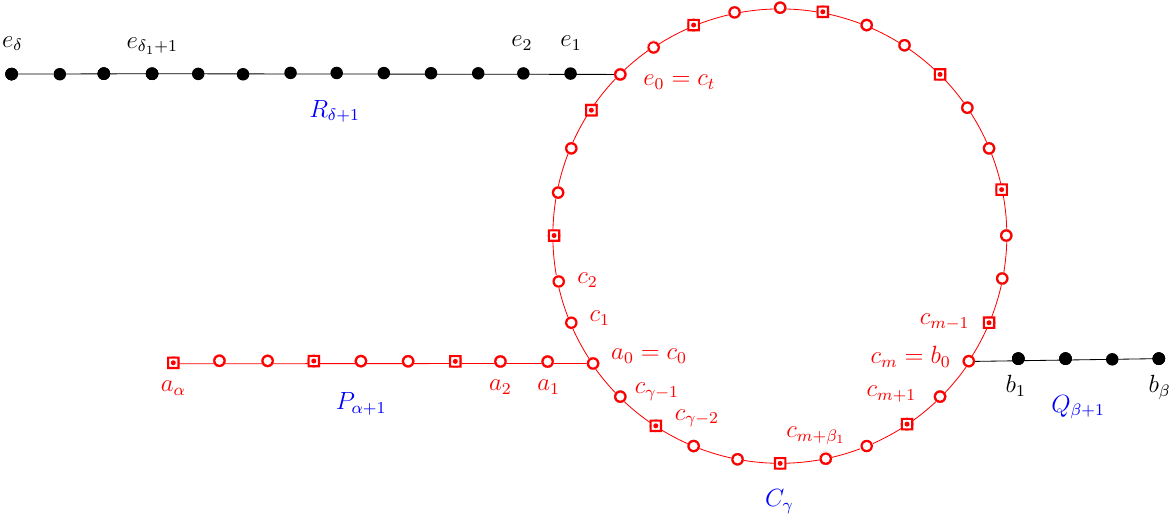}
 \caption{The circles and squares  represent the subgraph $P_{\alpha+1}\cup C_\gamma$  and the squares represent the  set $M'_\gamma(c_0,\alpha)$ in this figure.}
    \label{fig:multipacking_choice_2}
\end{figure}

\smallskip
\noindent \textbf{Finding a multipacking of  $\bm{H_\gamma(c_0,\alpha,c_t,\delta,c_m,\beta)} $ according to Choice-2 :}\label{subec:multipacking_technique_2}
\medskip

Here we find a path and a cycle of  $H$ so that the set of every third vertex (with some exceptions) 
on these subgraphs provide a multipacking of $H$. Consider the path $P_{\alpha+1}$ and the cycle $C_\gamma$. Now we choose every third vertex (with some exceptions) from these subgraphs  to construct a set, called $M'_\gamma(c_0,\alpha)$ (See Fig. \ref{fig:multipacking_choice_2}).  Formally we can define,  $M'_\gamma(c_0,\alpha)=\{a_i:0\leq i\leq \alpha, i\equiv 0 \text{ (mod $3$)}\}\cup \{c_i:0\leq i\leq \gamma-1,i\equiv 0 \text{ (mod $3$)}\}\setminus \{c_0\}$. 

Similarly, for  the path $Q_{\beta+1}$ and the cycle $C_\gamma$, we define the set $M'_\gamma(c_m,\beta)=\{b_i:0\leq i\leq \beta, i\equiv 0 \text{ (mod $3$)}\}\cup \{c_i:0\leq i\leq \gamma-1,i\equiv m \text{ (mod $3$)}\}\setminus \{c_m\}$. 
These sets will be a multipacking of $G$ under the conditions stated in the following lemma.

\begin{lemma} 
\label{lem:multipacking_delta=0}
     Let $G$ be a cactus and $H_\gamma(c_0,\alpha,c_t,\delta,c_m,\beta)$  be a subgraph of $G$. Then $M'_\gamma(c_0,\alpha)$ is a multipacking of $G$ of size at least $\big\lfloor\frac{\gamma}{3}\big\rfloor+\big\lfloor\frac{\alpha}{3}\big\rfloor-1$ and $M'_\gamma(c_m,\beta)$ is a multipacking of $G$ of size at least $\big\lfloor\frac{\gamma}{3}\big\rfloor+\big\lfloor\frac{\beta}{3}\big\rfloor-1$.
\end{lemma}

\begin{proof}  Let $H=H_\gamma(c_0,\alpha,c_t,\delta,c_m,\beta)$,  $M'=M'_\gamma(c_0,\alpha)$.  From the definition of $M'$,   $|N_r(v)\cap M'|\leq r$ for all $v\in V(H)$ and $r\geq 1$. Hence $M'$ is a multipacking of $H$ size at least 
  $\big\lfloor\frac{\gamma}{3}\big\rfloor+\big\lfloor\frac{\alpha}{3}\big\rfloor-1$. Therefore, $M'$ is a multipacking of $G$ by Lemma \ref{lem:multipacking_subgraph}. By the similar reason $M'_\gamma(c_m,\beta)$ is a multipacking of $G$ of size at least $\big\lfloor\frac{\gamma}{3}\big\rfloor+\big\lfloor\frac{\beta}{3}\big\rfloor-1$. 
\end{proof}

\begin{figure}[htp]
    \centering
    \includegraphics[  height=6.5cm]{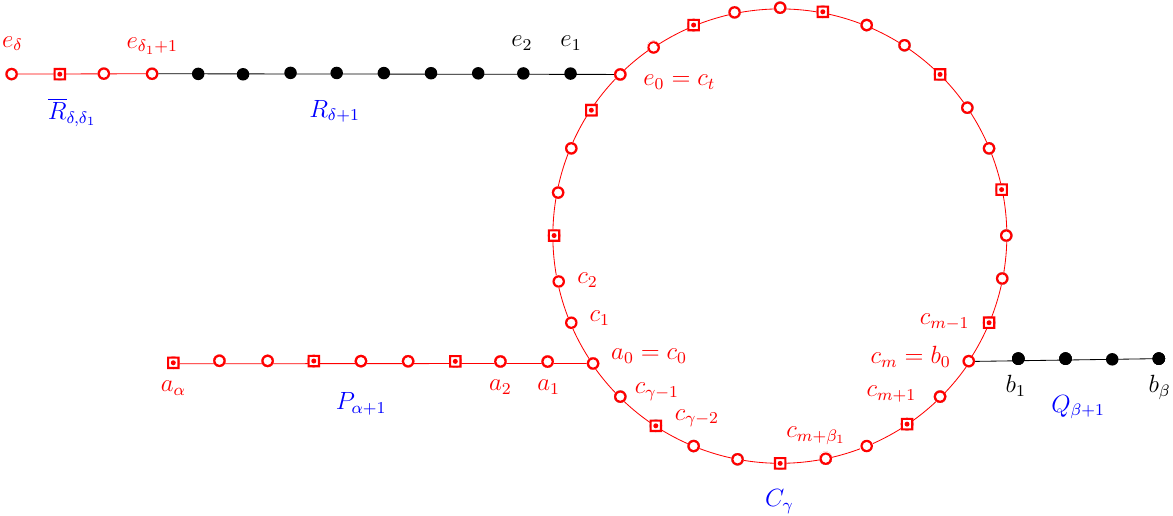}
 \caption{The circles and squares  represent the subgraph $P_{\alpha+1}\cup \overline{R}_{\delta,\delta_1}\cup C_\gamma$   and the squares represent the  set $M'_\gamma(c_0,\alpha,c_t,\delta,\delta_1)$ in this figure.}
    \label{fig:m2_11}
\end{figure}

\medskip
\noindent\textbf{Finding a multipacking in  $\bm{H_\gamma(c_0,\alpha,c_t,\delta,c_m,\beta)} $ according to Choice-3 :}\label{subec:multipacking_technique_3}
\smallskip

Next, we discuss a more general choice, similar to the last one. In the last choice, we did not use the path $R_{\delta+1}$ to construct a multipacking. Here we are going to use $R_{\delta+1}$ also when it is there in $G$. Let   $\overline{R}_{\delta,\delta_1}=(e_{\delta_1+1},e_{\delta_1+2},\dots,e_{\delta})$ which is a path and a part of the path $R_{\delta+1}$. 
Now we choose every third vertex (with some exceptions) from the subgraphs $P_{\alpha+1}$, $\overline{R}_{\delta,\delta_1}$ and $C_\gamma$   to construct a set, called $M'_\gamma(c_0,\alpha,c_t,\delta,\delta_1)$(See Fig. \ref{fig:m2_11}). Formally we can define,  $M'_\gamma(c_0,\alpha,c_t,\delta,\delta_1)=\{a_i:0\leq i\leq \alpha, i\equiv 0 \text{ (mod $3$)}\}\cup \{e_i:\delta_1+2\leq i\leq \delta,i\equiv \delta_1+1 \text{ (mod $3$)}\}\cup \{c_i:0\leq i\leq \gamma-1,i\equiv 0 \text{ (mod $3$)}\}\setminus \{c_0\}$.

Similarly, if we consider the subgraphs $Q_{\beta+1}$, $\overline{R}_{\delta,\delta_1}$ and $C_\gamma$, we can define,  $M'_\gamma(c_m,\beta,$ $ c_t,\delta,\delta_1)=\{b_i:0\leq i\leq \beta, i\equiv 0 \text{ (mod $3$)}\}\cup \{e_i:\delta_1+2\leq i\leq \delta,i\equiv \delta_1+1 \text{ (mod $3$)}\}\cup \{c_i:0\leq i\leq \gamma-1,i\equiv m \text{ (mod $3$)}\}\setminus \{c_m\}$.

These sets will be a multipacking of $G$ under some conditions, that has been stated in the following lemma.

\begin{lemma}  
\label{lem:delta1}
     Let $G$ be a cactus and $H_\gamma(c_0,\alpha,c_t,\delta,c_m,\beta)$  be a subgraph of $G$.  If $\delta_1=\big\lfloor\frac{\gamma}{2}\big\rfloor-d(c_0,c_t)$ and  $ \delta\geq \delta_1$, then $M'_\gamma(c_0,\alpha,c_t,\delta,\delta_1)$ is a multipacking of $G$ of size at least $\big\lfloor\frac{\gamma}{3}\big\rfloor+\big\lfloor\frac{\alpha}{3}\big\rfloor+ \big\lfloor\frac{\delta-\delta_1}{3}\big\rfloor-1$. Moreover, if $\delta_2=\big\lfloor\frac{\gamma}{2}\big\rfloor-d(c_m,c_t)$ and $ \delta\geq \delta_2$, then $M'_\gamma(c_m,\beta,c_t,\delta,\delta_2)$ is a multipacking of $G$ of size at least $\big\lfloor\frac{\gamma}{3}\big\rfloor+\big\lfloor\frac{\beta}{3}\big\rfloor+ \big\lfloor\frac{\delta-\delta_2}{3}\big\rfloor-1$.
\end{lemma}

\begin{proof} Let $H=H_\gamma(c_0,\alpha,c_t,\delta,c_m,\beta)$,  $M'=M'_\gamma(c_0,\alpha,c_t,\delta,\delta_1)$.

Here $V(C_\gamma)=\{c_i:0\leq i\leq \gamma-1\}$. 
Let $v=c_{x_1}$ for some $x_1$ where $0\leq x_1\leq \gamma-1$. Let  $d_1=d(c_0,v)$ and $d_2=d(c_t,v)$. Since $c_0,c_t\in V(C_\gamma)$,  $d(c_0,c_t)\leq diam(C_\gamma)= \big\lfloor\frac{\gamma}{2}\big\rfloor$. Therefore $\big\lfloor\frac{\gamma}{2}\big\rfloor- d(c_0,c_t)\geq 0\implies \delta_1\geq 0$. 

\vspace{0.22cm}

 \noindent \textit{\textbf{Case 1: }}
    $1\leq r \leq \max\{d_1,d_2+\delta_1\}$.

In that case, $|N_r(v)\cap M'|\leq r$ for $v\in V(C_\gamma)$.

\vspace{0.22cm}
   
 \noindent \textit{\textbf{Case 2: }}  $ \max\{d_1,d_2+\delta_1\}<r$. 
 
 $|N_r(v)\cap M'|\leq \frac{1}{3}[r+1+r-d_1+r-\delta_1]=\frac{1}{3}[3r+1-d_1-\delta_1]\leq r$.

 \vspace{0.22cm}

 Suppose $v\in V(P_{\alpha+1})$, then any path that joins $v$ with a vertex in $V(C_\gamma)\cup V(Q_{\beta+1})\cup V(R_{\delta+1})$ passes through $a_0$, otherwise $G$ cannot be a cactus. By Observation \ref{obs:isometric_graph} we know that $H_\gamma(c_0,\alpha,c_t,\delta,c_m,\beta)$ is an isometric subgraph of $G$.  Therefore $|N_r(v)\cap M'|\leq r$ for all  $r\geq 1$. Similarly we can show that, if $v\in V(Q_{\beta+1})$ or $v \in V(R_{\delta+1})$, then $|N_r(v)\cap M'|\leq r$ for all  $r\geq 1$. Therefore $M'$ is a multipacking of $H$. So, $M'$ is a multipacking of $G$ by Lemma \ref{lem:multipacking_subgraph}. $M'_\gamma(c_0,\alpha,c_t,\delta,\delta_1)$ is a multipacking of $G$ of size at least $\big\lfloor\frac{\gamma}{3}\big\rfloor+\big\lfloor\frac{\alpha}{3}\big\rfloor+ \big\lfloor\frac{\delta-\delta_1}{3}\big\rfloor-1$.
 
 Similarly, we can show that, if $\delta_2=\big\lfloor\frac{\gamma}{2}\big\rfloor-d(c_m,c_t)$ and $ \delta\geq \delta_2$, then $M'_\gamma(c_m,\beta,c_t,$ $\delta,\delta_2)$ is a multipacking of $G$ of size at least $\big\lfloor\frac{\gamma}{3}\big\rfloor+\big\lfloor\frac{\beta}{3}\big\rfloor+ \big\lfloor\frac{\delta-\delta_2}{3}\big\rfloor-1$.  
\end{proof}

Now we are ready to prove Lemma  \ref{lem:multipacking_radius_relation}. Here is a small observation before we start proving the lemma. We use this observation in the proof.

\begin{observation}
\label{obs:r/3} If $r$ is a positive integer, then 
    $\big\lfloor\frac{r}{3}\big\rfloor+\big\lfloor\frac{r-1}{3}\big\rfloor \geq \big\lfloor\frac{2r-1}{3}\big\rfloor-1$,   $\big\lfloor\frac{r}{3}\big\rfloor\geq \frac{r}{3}-\frac{2}{3}$,  and $\big\lfloor\frac{r}{2}\big\rfloor+\big\lceil\frac{r}{2}\big\rceil=r$.



\end{observation}

\begin{lemma} 
    \label{lem:multipacking_radius_relation}
  Let $G$ be a cactus with radius $\rad(G)$, then  $\MP(G)\geq \frac{2}{3}\rad(G)-\frac{11}{3}$.
\end{lemma}

\begin{proofsketch}     First, we present a very brief sketch of the proof for Lemma \ref{lem:multipacking_radius_relation}. Let $\rad(G)=r$ and $c$ be a center of $G$. If $r=0$ or $1$, then $\MP(G)=1$. Therefore, for $r\leq 1$, we have $\MP(G)\geq \frac{2}{3}\rad(G)$. Now assume $r\geq 2$. Since $G$ has radius $r$, there is an isometric path $P$ in $G$ whose one endpoint  is $c$ and $l(P)=r$. Let $Q$ be a  largest isometric path in $G$ whose one endpoint  is $c$ and  $V(P)\cap V(Q)=\{c\}$. If $l(Q)=r'$, then $ r-1 \leq r'\leq  r$  by Lemma \ref{lem:disjoint_radial_path}.
    Let $P=(v_0,v_1, \dots , v_{r})$ and  $Q=(w_0,w_1, \dots , w_{r'})$ where $v_0=w_0=c$. From Observation \ref{obs:joining_path_atmost_1}, we know that $|X_{P,Q}|\leq 1$. 

     If  $|X_{P,Q}|=0$, we take the path $P\cup Q$ which is an isometric path of length $r+r'$ and choose every third vertex to the path to construct a multipacking of size at least $\MP(G)\geq \big\lceil\frac{2}{3}\rad(G)\big\rceil$ by Lemma \ref{lem:isometric_path}.

     Suppose $|X_{P,Q}|=1$. We have already discussed in this section that we can find a subgraph $H_\gamma(c_0,\alpha,c_t,\delta,$ $c_m,\beta)$ in $G$ and we use Lemma \ref{lem:gamma/2}, Lemma \ref{lem:multipacking_delta=0} and Lemma \ref{lem:delta1} under several cases to construct a multipacking of size at least $ \frac{2}{3}\rad(G)-\frac{11}{3}$.   
\end{proofsketch}

\begin{proof}[Proof of Lemma \ref{lem:multipacking_radius_relation}] Let $\rad(G)=r$ and $c$ be a center of $G$. If $r=0$ or $1$, then $\MP(G)=1$. Therefore, for $r\leq 1$, we have $\MP(G)\geq \frac{2}{3}\rad(G)$. Now assume $r\geq 2$. Since $G$ has radius $r$, there is an isometric path $P$ in $G$ whose one endpoint  is $c$ and $l(P)=r$. Let $Q$ be a  largest isometric path in $G$ whose one endpoint  is $c$ and  $V(P)\cap V(Q)=\{c\}$. If $l(Q)=r'$, then $ r-1 \leq r'\leq  r$  by Lemma \ref{lem:disjoint_radial_path}.
    Let $P=(v_0,v_1, \dots , v_{r})$ and  $Q=(w_0,w_1, \dots , w_{r'})$ where $v_0=w_0=c$. From Observation \ref{obs:joining_path_atmost_1}, we know that $|X_{P,Q}|\leq 1$.   

\vspace{0.2cm}
\noindent
\textbf{Claim \ref{lem:multipacking_radius_relation}.1. } If  $|X_{P,Q}|=0$, then $\MP(G)\geq \big\lceil\frac{2}{3}\rad(G)\big\rceil$.

\begin{claimproof} Since  $|X_{P,Q}|=0$,  any path in $G$ that joins a vertex of $P$ with a vertex of $Q$ always passes through $c$. Therefore, $P\cup Q$ is an isometric path of length $r+r'$ and $|V(P\cup Q)|=r+r'+1$. By Lemma \ref{lem:isometric_path}, there is a multipacking in $G$ of size $\big\lceil\frac{r+r'+1}{3}\big\rceil$. Since $r'\geq r-1$, $\big\lceil\frac{r+r'+1}{3}\big\rceil \geq \big\lceil\frac{2r}{3}\big\rceil$.  Therefore, $\MP(G)\geq \big\lceil\frac{2r}{3}\big\rceil$.    
\end{claimproof} 

\medskip

Suppose $|X_{P,Q}|=1$. Let $X_{P,Q}=\{(v_i,w_j)\}$. Then $|X_{P,Q}(v_i,w_j)|=1$ by Observation \ref{obs:joining_path_atmost_1}. Let  $F_1\in X_{P,Q}(v_i,w_j)$ and $F_2=(v_i,v_{i-1},v_{i-2},\dots,v_1,v_0,w_1,w_2,\dots,$ $w_{j-1},w_j)$. Therefore, $F_1$ and $F_2$ form a cycle $F_1\cup F_2$ of length  $ l(F_1)+l(F_2)$. Note that $l(F_1)\geq 1$, since $v_i\neq w_j$. Let $\gamma= l(F_1)+l(F_2)$ and  $C_\gamma=(c_0,c_1,c_2,\dots,c_{\gamma-2},$ $c_{\gamma-1},c_0) $ be a cycle of length $\gamma$.  Therefore,  $C_\gamma$ is  isomorphic to $ F_1\cup F_2$. For simplicity, we assume that $C_\gamma = F_1\cup F_2$. So, we can  assume that  $F_2=(c_0,c_1,c_2,\dots,$ $c_{m})$ and $F_1=(c_m,c_{m+1},\dots,c_{\gamma-1}, c_0)$.  
 Since $P$ and $Q$ are isometric paths, therefore $P'=(v_i,v_{i+1},\dots,v_{r})$ and $Q'=(w_j,w_{j+1},\dots,w_{r'})$ are also isometric paths in $G$.  Therefore $C_\gamma\cup P'\cup Q'$ can be represented as 
 $H_\gamma(c_0,\alpha,c_t,0,c_m,\beta)$ or $ H_\gamma(c_0,\alpha,c_m,\beta)$, where $\alpha =l(P')$ and $\beta=l(Q')$. So, $C_\gamma\cup P'\cup Q'$ is an isometric subgraph of $G$ by Observation \ref{obs:isometric_graph}. Let $H=C_\gamma\cup P'\cup Q'$. For simplicity, we can assume that $P'=P_{\alpha+1}=(a_0,a_1,\dots,a_{\alpha+1})$ and $Q'=Q_{\beta+1}=(b_0,b_1,\dots,b_{\alpha+1})$. (See. Fig. \ref{fig:bfs3} and Fig. \ref{fig:bfs1}) 
 
 \vspace{0.2cm}
\noindent
\textbf{Claim \ref{lem:multipacking_radius_relation}.2. }  If  $|X_{P,Q}|=1$ and $l(F_1)\geq  l(F_2)$, then $\MP(G)\geq \big\lceil\frac{2}{3}\rad(G)\big\rceil$.

\begin{claimproof}
Since $G$ is a cactus and $l(F_1)\geq  l(F_2)$,  
$P\cup Q$ is an isometric path in $G$ by Observation \ref{obs:isometric_graph_path_F_1}.  Note that  $l(P\cup Q)=r+r'$ and $|V(P\cup Q)|=r+r'+1$. By Lemma \ref{lem:isometric_path}, we can say that, there is a multipacking in $G$ of size $\big\lceil\frac{r+r'+1}{3}\big\rceil$. Now $r'\geq r-1$ $\implies$ $\big\lceil\frac{r+r'+1}{3}\big\rceil \geq \big\lceil\frac{2r}{3}\big\rceil$.  Therefore, $\MP(G)\geq \big\lceil\frac{2r}{3}\big\rceil$.  
\end{claimproof}

 \medskip
 
 Now assume $l(F_1)<  l(F_2)$. Therefore $F_1\cup P'\cup Q'$ is an isometric path by Observation \ref{obs:isometric_graph_path_F_1}. We know $l(F_2)=m$. Let $l(F_1)=x$   and $g=m+\big\lfloor\frac{x}{2}\big\rfloor$. Therefore $c_g$ is a middle point of the path $F_1$. Let $S_r=\{u\in V(G):d(c_g,u)=r\}$.  Here  $S_r\neq \phi$, since $\rad(G)=r$. We know that $c$ is a center of $G$ and  $c \in V(C_\gamma)$, more precisely $c\in V(F_2)$. Therefore $c=c_k$ for some $k\in \{0,1,2,\dots,m\}$. Let $F_2^1=(c_0,c_1,\dots,c_k)$ and $F_2^2=(c_k,c_{k+1},\dots,c_m)$. Therefore $F_2=F_2^1\cup F_2^2$. Let  $l(F_2^1)=y$ and $l(F_2^2)=z$. Therefore, $F_1\cup F_2^1\cup F_2^2=C_\gamma$ and $x+y+z=\gamma$. Note that  $d(c_g,c_m)=d_H(c_g,c_m)=\big\lfloor\frac{x}{2}\big\rfloor$ and $d(c_g,c_0)=d_H(c_g,c_0)=x-\big\lfloor\frac{x}{2}\big\rfloor=\big\lceil\frac{x}{2}\big\rceil$.  We know that $l(P')=\alpha $ and $l(Q')=\beta$. Therefore $\alpha +y=l(P')+l(F^1_2)=l(P)=r$ and $\beta+z=l(Q')+l(F^2_2)=l(Q)=r'$. From now we use all these notations in the upcoming results.

Now we want to show that,  if   $l(F_1)<  l(F_2)$, then $x$, $y$ and $z$ are upper bounded by $\big\lfloor\frac{\gamma}{2}\big\rfloor$.  
 $l(F_1)<  l(F_2)\implies x<y+z\implies x<\frac{x+y+z}{2}\implies x<\frac{\gamma}{2} \implies x\leq \big\lfloor\frac{\gamma}{2}\big\rfloor$.  Since $P$ is an isometric path, $F_2^1$ is a shortest path joining $c_0$ and $c$. Note that, the path  $ F_2^2\cup F_1$  also  joins $c_0$ and $c$. Therefore, $l(F_2^1)\leq l(F_2^2\cup F_1)\implies l(F_2^1)\leq l(F_2^2)+l(F_1)\implies y\leq z+x \implies y\leq \frac{x+y+z}{2}\implies y\leq \big\lfloor\frac{\gamma}{2}\big\rfloor$. Similarly, we can show that  $z\leq \big\lfloor\frac{\gamma}{2}\big\rfloor$, since $F_2^2$ is a shortest path joining $c_0$ and $c$. 
 Therefore, if $l(F_1)<  l(F_2)$ , $\max\{x,y,z\}\leq \big\lfloor\frac{\gamma}{2}\big\rfloor$. 

\vspace{0.2cm}
\noindent
\textbf{Claim \ref{lem:multipacking_radius_relation}.3. }  If  $|X_{P,Q}|=1$, $l(F_1)<  l(F_2)$ and $S_r\cap P'\neq \phi $, then $\MP(G)\geq \big\lfloor\frac{2}{3}\rad(G)-\frac{1}{3}\big\rfloor-3$.

\begin{claimproof}
   Let $u\in S_r\cap P'$. Let $\alpha_1=x-1$ and  $\beta_1=z-1$.  Since $F_1\cup P' \cup Q' $ is an isometric path of $G$,   $\alpha+\alpha_1+1=x+\alpha =  d(c_m,v_r)\geq d(c_g,u)=r $. Here $\beta+\beta_1+1=z+\beta=r'\geq r-1$. We have $\alpha_1=x-1\leq \big\lfloor\frac{\gamma}{2}\big\rfloor-1$ and $\beta_1=z-1\leq \big\lfloor\frac{\gamma}{2}\big\rfloor-1$, since $\max\{x,y,z\}\leq \big\lfloor\frac{\gamma}{2}\big\rfloor$. We have shown that $H$ can be represented as 
 $ H_\gamma(c_0,\alpha,c_t,0,c_m,\beta) $. Therefore, there is a multipacking of $G$ of size at least 
  $ \big\lfloor\frac{\alpha+\alpha_1+1}{3}\big\rfloor+\big\lfloor\frac{\beta+\beta_1+1}{3}\big\rfloor -2$ by Lemma \ref{lem:gamma/2}.  
 Here $  \big\lfloor\frac{\alpha+\alpha_1+1}{3}\big\rfloor+\big\lfloor\frac{\beta+\beta_1+1}{3}\big\rfloor -2\geq \big\lfloor\frac{r}{3}\big\rfloor+\big\lfloor\frac{r-1}{3}\big\rfloor -2\geq \big\lfloor\frac{2r-1}{3}\big\rfloor-3$ by Observation \ref{obs:r/3}. 
\end{claimproof}

\vspace{0.2cm}
\noindent
\textbf{Claim \ref{lem:multipacking_radius_relation}.4. }  If  $|X_{P,Q}|=1$, $l(F_1)<  l(F_2)$ and $S_r\cap Q'\neq \phi $, then $\MP(G)\geq 2\big\lfloor\frac{\rad(G)}{3}\big\rfloor-2$.

\begin{claimproof}
Let $u\in S_r\cap Q'$. Let $\alpha_1=y-1$ and   $\beta_1=x-1$.  Since $F_1\cup P' \cup Q' $ is an isometric path of $G$, $\beta+\beta_1+1=x+\beta=d(c_0,w_{r'})\geq d(c_g,u)=r $. Moreover, $\alpha+\alpha_1+1=y+\alpha =  r$.  We have $\alpha_1=y-1\leq \big\lfloor\frac{\gamma}{2}\big\rfloor-1$ and $\beta_1=x-1\leq \big\lfloor\frac{\gamma}{2}\big\rfloor-1$, since $\max\{x,y,z\}\leq \big\lfloor\frac{\gamma}{2}\big\rfloor$. We have shown that $H$ can be represented as 
 $ H_\gamma(c_0,\alpha,c_t,0,c_m,\beta) $. Therefore,  there is a multipacking of $G$ of size at least 
  $ \big\lfloor\frac{\alpha+\alpha_1+1}{3}\big\rfloor+\big\lfloor\frac{\beta+\beta_1+1}{3}\big\rfloor -2$ by Lemma \ref{lem:gamma/2}. 
 Here $  \big\lfloor\frac{\alpha+\alpha_1+1}{3}\big\rfloor+\big\lfloor\frac{\beta+\beta_1+1}{3}\big\rfloor -2\geq \big\lfloor\frac{r}{3}\big\rfloor+\big\lfloor\frac{r}{3}\big\rfloor -2\geq 2\big\lfloor\frac{r}{3}\big\rfloor-2$.  
\end{claimproof}

\vspace{0.2cm}
\noindent
\textbf{Claim \ref{lem:multipacking_radius_relation}.5. } If  $|X_{P,Q}|=1$, $l(F_1)<  l(F_2)$ and $S_r\cap C_\gamma \neq \phi $, then $\MP(G)\geq \big\lceil\frac{2}{3}\rad(G)\big\rceil$.

\begin{claimproof}
    We know that $ H_\gamma(c_0,0,c_t,0,c_m,0)= C_\gamma$.  Therefore $C_\gamma$ is an isometric subgraph of $G$ by Observation \ref{obs:isometric_graph}. $S_r\cap C_\gamma \neq \phi \implies r\leq \big\lceil\frac{\gamma}{2}\big\rceil\implies 2r\leq \gamma$. Let $M=\{c_i:0\leq i\leq \gamma-1,i\equiv 0 \text{ (mod $3$)}\}$. Note that, $M$ is a multipacking of $C_\gamma$. Therefore, $M$ is a mutipacking of $G$ by Lemma \ref{lem:multipacking_subgraph}.  Here $|M|=\big\lceil\frac{\gamma}{3}\big\rceil\geq \big\lceil\frac{2r}{3}\big\rceil$, since $2r\leq \gamma$.   
\end{claimproof}

\vspace{0.2cm}
\noindent
\textbf{Claim \ref{lem:multipacking_radius_relation}.6. } If  $|X_{P,Q}|=1$, $l(F_1)<  l(F_2)$ and $S_r\cap V(H) =\phi $, then $\MP(G)\geq \frac{2}{3}\rad(G)-\frac{11}{3}$.

\begin{claimproof} Let $u\in S_r$. Therefore, $u\notin V(H)$. We know that $g=m+\big\lfloor\frac{x}{2}\big\rfloor$. Let $R$ be a shortest path joining $c_g$ and $u$. Therefore, $R$ is an isometric path of $G$. Let $R=(u_0,u_1,\dots,u_r)$ where $c_g=u_0$ and $u=u_r$. Suppose $h=\max\{i:u_i\in V(H)\}$.   Let $R'=(u_{h+1},u_{h+2},\dots,u_r)$. 

First, we show that  $u_h\notin V(F_1)$. Suppose $u_h\in V(F_1)$, so $u_h\in \{c_m,c_{m+1},\dots,$ $c_{\gamma-1},c_0\}$. Therefore, $u_h=c_{g'}$ for some $g'\in \{m,m+1,\dots,\gamma-1,0\}$. First consider $u_h\in \{c_{g+1},c_{g+2},\dots,$ $c_{\gamma-1},c_0\}$. We know that $c$ is a center of $G$ and $c=c_k$ for some $k\in \{0,1,2,\dots,m\}$. Suppose $(c_k,c_{k-1},\dots,c_0,c_{\gamma-1},\dots,c_{g'})$ is a shortest path joining $c_k(=c)$ and $c_{g'}(=u_h)$.  We have shown that $H$ is an isometric subgraph of $G$. Therefore, the distance between two vertices in $H$ is the distance in $G$.
Since $S_r\cap V(H)= \phi$, we have $d(c_g,v_r)<r$. Therefore,  $d(c_g,v_r)<r=d(c_g,u)\implies d(c_g,v_r)<d(c_g,u) \implies d(c_g,u_h)+d(u_h,c_0)+d(c_0,v_r)<d(c_g,u_h)+d(u_h,u)\implies d(u_h,c_0)+d(c_0,v_r)<d(u_h,u)\implies d(c_0,v_r)<d(u_h,u)$.   Since $c$ is a center of $G$ and $S_r\cap V(H)= \phi$, we have $r\geq d(c,u)=d(c,c_0)+d(c_0,u_h)+d(u_h,u)>d(c,c_0)+d(c_0,u_h)+d(c_0,v_r)=d(c,v_r)+d(c_0,u_h)=r+d(c_0,u_h)\implies d(c_0,u_h)<0$. This is a contradiction. Now assume $(c_k,c_{k+1},\dots,c_m,c_{m+1},\dots,c_{g'})$ is a shortest path joining $c$ and $u_h$. Since $c$ is a center of $G$ and $S_r\cap V(H)= \phi$, we have $r\geq d(c,u)=d(c,c_g)+d(c_g,u)=d(c,c_g)+r\implies d(c,c_g)=0\implies c=c_g$. Therefore $r=d(c,v_{r})=d(c_g,v_r)\implies v_r\in S_r\cap V(H)$, which is a contradiction. 
Therefore, $u_h\notin \{c_{g+1},c_{g+2},\dots,c_{\gamma-1},c_0\}$. Similarly we can show that $u_h\notin \{c_{m},c_{m+1},\dots,c_{g-1},c_g\}$. So, $u_h\notin V(F_1)$.

If $u_h\in V(P')$, then $d(c_g,u)=r>d(c_g,v_r)\implies d(c_g,u_h)+d(u,u_h)>d(c_g,u_h)+d(u_h,v_r)\implies d(u,u_h)>d(u_h,v_r)\implies d(u,u_h)+d(c,u_h)>d(u_h,v_r)+d(c,u_h)\implies d(c,u)>d(c,v_r)\implies d(c,u)>r$, which is a contradiction, since $c$ is a center of the graph $G$ having radius $r$. Therefore, $u_h\notin V(P')$. Similarly, we can show that $u_h\notin V(Q')$.



Therefore $u_h\in V(C_\gamma)\setminus V(F_1)= \{c_1,c_2,\dots,c_{m-1}\}$.  Since $G$ is a cactus, therefore every edge belongs to almost one cycle. This implies that $u_i\in V(C_\gamma)$ for all $0\leq i\leq h$.  Let $u_h=c_t$. 

Suppose $x\geq \alpha $, then $x+y+z+\beta\geq \alpha +y+z+\beta\geq r+r'\geq 2r-1$.  By Lemma \ref{lem:multipacking_delta=0},  $M'_\gamma(c_m,\beta)$ is a multipacking of $G$ of size at least $\big\lfloor\frac{\gamma}{3}\big\rfloor+\big\lfloor\frac{\beta}{3}\big\rfloor-1$. Therefore  $\big\lfloor\frac{\gamma}{3}\big\rfloor+\big\lfloor\frac{\beta}{3}\big\rfloor-1\geq \frac{\gamma}{3}-\frac{2}{3}+\frac{\beta}{3}-\frac{2}{3}-1=\frac{x+y+z+\beta}{3}-\frac{7}{3}\geq \frac{2r-1}{3}-\frac{7}{3}$. 

Suppose $x\geq \beta$, then $x+y+z+\alpha \geq \alpha +y+z+\beta\geq r+r'\geq 2r-1$.  By Lemma \ref{lem:multipacking_delta=0},  $M'_\gamma(c_0,\alpha )$ is a multipacking of $G$ of size at least $\big\lfloor\frac{\gamma}{3}\big\rfloor+\big\lfloor\frac{\alpha }{3}\big\rfloor-1$ (See Fig. \ref{fig:multipacking_choice_2}). Therefore  $\big\lfloor\frac{\gamma}{3}\big\rfloor+\big\lfloor\frac{\alpha }{3}\big\rfloor-1\geq \frac{\gamma}{3}-\frac{2}{3}+\frac{\alpha }{3}-\frac{2}{3}-1=\frac{x+y+z+\beta}{3}-\frac{7}{3}\geq \frac{2r-1}{3}-\frac{7}{3}$. 

Now assume $x<\min\{\alpha ,\beta\}$.

Note that, $S_r\cap V(H)=\phi\implies S_r\cap V(C_\gamma)=\phi$. Therefore, there is no vertex on the cycle $C_\gamma$ which is at  distance $r$ from the vertex $c_g$. Therefore, $r>\big\lfloor\frac{\gamma}{2}\big\rfloor$.

Now we split the remainder of the proof  into two cases.

\vspace{0.22cm} 

 \noindent \textit{\textbf{Case 1: }} $ \big\lfloor\frac{\gamma}{2}\big\rfloor<r\leq \big\lfloor\frac{\gamma}{2}\big\rfloor+\big\lfloor\frac{x}{2}\big\rfloor$.

 Consider the set $M'_\gamma(c_0,\alpha )$. This a multipacking of $G$ of size $\big\lfloor\frac{\gamma}{3}\big\rfloor+\big\lfloor\frac{\alpha }{3}\big\rfloor-1$ by Lemma \ref{lem:multipacking_delta=0}. Now $\big\lfloor\frac{\gamma}{3}\big\rfloor+\big\lfloor\frac{\alpha }{3}\big\rfloor-1\geq
 \big\lfloor\frac{\gamma}{3}\big\rfloor+\big\lfloor\frac{x}{3}\big\rfloor-1\geq
 \frac{\gamma}{3}-\frac{2}{3}+\frac{x}{3}-\frac{2}{3}-1\geq \frac{2}{3}.\big(\frac{\gamma}{2}+\frac{x}{2}\big)-\frac{7}{3}\geq \frac{2}{3}r-\frac{7}{3}$.

\vspace{0.22cm} 

 \noindent \textit{\textbf{Case 2: }} $ \big\lfloor\frac{\gamma}{2}\big\rfloor+\big\lfloor\frac{x}{2}\big\rfloor<r$.

Now consider  
 $\delta=r-d(c_g,c_t)$, $\delta_1=\big\lfloor\frac{\gamma}{2}\big\rfloor-d(c_0,c_t)$ and $\delta_2=\big\lfloor\frac{\gamma}{2}\big\rfloor-d(c_m,c_t)$.  Since $c_0,c_t\in V(C_\gamma)$,  $d(c_0,c_t)\leq diam(C_\gamma)= \big\lfloor\frac{\gamma}{2}\big\rfloor$. Therefore $\big\lfloor\frac{\gamma}{2}\big\rfloor- d(c_0,c_t)\geq 0\implies \delta_1\geq 0$. Similarly, we can show that $\delta_2\geq 0$. Now $\delta-\delta_1=r-d(c_g,c_t)-\big\lfloor\frac{\gamma}{2}\big\rfloor+d(c_0,c_t)\geq r-d(c_g,c_0)-d(c_0,c_t)-\big\lfloor\frac{\gamma}{2}\big\rfloor+d(c_0,c_t)\geq r-\big\lceil\frac{x}{2}\big\rceil-\big\lfloor\frac{\gamma}{2}\big\rfloor\geq 0$. Therefore $\delta \geq \delta_1$ and $\delta - \delta_1\geq r-\big\lceil\frac{x}{2}\big\rceil-\big\lfloor\frac{\gamma}{2}\big\rfloor\geq 0$. Similarly, we can show that $\delta\geq \delta_2$ and $\delta - \delta_2\geq r-\big\lceil\frac{x}{2}\big\rceil-\big\lfloor\frac{\gamma}{2}\big\rfloor\geq 0$.  Now we are ready to use Lemma \ref{lem:delta1} to find multipacking in $G$.


First assume that, $z\geq y$. 
By Lemma \ref{lem:delta1},   $M'_\gamma(c_0,\alpha ,c_t,\delta,\delta_1)$ is a multipacking of $G$ of size at least $\big\lfloor\frac{\gamma}{3}\big\rfloor+\big\lfloor\frac{\alpha }{3}\big\rfloor+ \big\lfloor\frac{\delta-\delta_1}{3}\big\rfloor-1$.  Now, $\big\lfloor\frac{\gamma}{3}\big\rfloor+\big\lfloor\frac{\alpha }{3}\big\rfloor+ \big\lfloor\frac{\delta-\delta_1}{3}\big\rfloor-1\geq
\frac{\gamma}{3}-\frac{2}{3}+\frac{\alpha }{3}-\frac{2}{3}+\frac{\delta-\delta_1}{3}-\frac{2}{3}-1=
\frac{\gamma}{3}+\frac{\alpha }{3}+\frac{1}{3}.\big(r-\big\lceil\frac{x}{2}\big\rceil-\big\lfloor\frac{\gamma}{2}\big\rfloor\big)-3   
\geq  \frac{\gamma}{3}+\frac{\alpha }{3}+\frac{1}{3}.\big(r-\frac{x}{2}-1-\frac{\gamma}{2}\big)-3 =
\frac{1}{3}\big(r+\frac{\gamma}{2}-\frac{x}{2}+\alpha \big) -\frac{10}{3}\geq 
\frac{1}{3}\big(r+\frac{\gamma}{2}-\frac{x}{2}+r-y\big) -\frac{10}{3}\geq
\frac{1}{3}\big(2r+\frac{x+y+z}{2}-\frac{x}{2}-y\big) -\frac{10}{3}\geq
\frac{1}{3}\big(2r+\frac{z-y}{2}\big) -\frac{10}{3}\geq \frac{2}{3}r-\frac{10}{3}$.

Suppose $z<y$. By Lemma \ref{lem:delta1},   $M'_\gamma(c_m,\beta,c_t,\delta,\delta_2)$ is a multipacking of $G$ of size at least $\big\lfloor\frac{\gamma}{3}\big\rfloor+\big\lfloor\frac{\beta}{3}\big\rfloor+ \big\lfloor\frac{\delta-\delta_2}{3}\big\rfloor-1$.  Now, $\big\lfloor\frac{\gamma}{3}\big\rfloor+\big\lfloor\frac{\beta}{3}\big\rfloor+ \big\lfloor\frac{\delta-\delta_2}{3}\big\rfloor-1\geq
\frac{\gamma}{3}-\frac{2}{3}+\frac{\beta}{3}-\frac{2}{3}+\frac{\delta-\delta_2}{3}-\frac{2}{3}-1=
\frac{\gamma}{3}+\frac{\beta}{3}+\frac{1}{3}.\big(r-\big\lceil\frac{x}{2}\big\rceil-\big\lfloor\frac{\gamma}{2}\big\rfloor\big)-3   
\geq  \frac{\gamma}{3}+\frac{\beta}{3}+\frac{1}{3}.\big(r-\frac{x}{2}-1-\frac{\gamma}{2}\big)-3 =
\frac{1}{3}\big(r+\frac{\gamma}{2}-\frac{x}{2}+\beta\big) -\frac{10}{3}\geq 
\frac{1}{3}\big(r+\frac{\gamma}{2}-\frac{x}{2}+r'-z\big) -\frac{10}{3}\geq
\frac{1}{3}\big(r+r-1+\frac{x+y+z}{2}-\frac{x}{2}-z\big) -\frac{10}{3}\geq
\frac{1}{3}\big(2r+\frac{y-z}{2}\big) -\frac{11}{3}\geq \frac{2}{3}r-\frac{11}{3}$. 
\end{claimproof}

In each case, $G$ has a multipacking of size at least $ \frac{2}{3}\rad(G)-\frac{11}{3}$. Therefore, $\MP(G)\geq \frac{2}{3}\rad(G)-\frac{11}{3}$.  
\end{proof}

\begin{theorem}[\cite{teshima2012broadcasts,erwin2001cost}]  \label{thm:mpGleqgammabG}  If $G$ is a connected graph of order at least 2  having radius $ \rad(G) $, multipacking number  $\MP(G) $, broadcast domination number $ \gamma_{b}(G) $ and domination number $\gamma(G)$, then $ \MP(G)\leq \gamma_{b}(G) \leq min\{\gamma (G),\rad(G)\}$.
\end{theorem}

\begin{proof}[Proof of Theorem \ref{thm:multipacking_broadcast_relation}]
    We have $\gamma_b(G)\leq \rad(G)$ from Theorem \ref{thm:mpGleqgammabG}. From Lemma 
\ref{lem:multipacking_radius_relation}, we get $\MP(G)\geq \frac{2}{3}\rad(G)-\frac{11}{3}$ . Therefore, $ \frac{2}{3}.\gamma_b(G)-\frac{11}{3}\leq \MP(G)\implies \gamma_b(G)\leq \frac{3}{2}\MP(G)+\frac{11}{2}$.
\end{proof}

\section{An approximation algorithm to find multipacking in cactus graphs}\label{sec:approximation_algorithm_to_find_Multipacking}

We provide the following algorithm to approximate multipacking of a cactus graph. This algorithm is a direct implementation of the proof of Theorem \ref{thm:multipacking_broadcast_relation}.

\smallskip
\noindent\textbf{Approximation algorithm: } Since $G$ is a cactus graph,  we can find a center $c$ and  an isometric path $P$ of length $r$ whose one endpoint is $c$ in $O(n)$-time, where $n$ is the number of vertices of $G$. After that, we can find another isometric path $Q$ having length $r-1$ or $r$ whose one endpoint is $c$ and $V(P)\cap V(Q)=\{c\}$.  Then the flow of the proof of 
Lemma  \ref{lem:multipacking_radius_relation} provides an $O(n)$-time algorithm to construct a multipacking of size at least $\frac{2}{3}\rad(G)-\frac{11}{3}$. Thus we can find  a multipacking of $G$ of size at least $\frac{2}{3}\rad(G)-\frac{11}{3}$ in $O(n)$-time.  Moreover,  $\MP(G)\geq \frac{2}{3}\rad(G)-\frac{11}{3}\geq \frac{2}{3}\MP(G)-\frac{11}{3}$ by Theorem \ref{thm:mpGleqgammabG}. This implies the existence of a $(\frac{3}{2}+o(1))$-factor approximation algorithm for
the multipacking problem on cactus graph. Therefore, we have the following.

\MultipackingAlgorithm*

\section{Unboundedness of the gap between broadcast domination and multipacking number of cactus and AT-free graphs}\label{sec:Unboundedness_of_the_gap_between_Broadcast_domination_and_Multipacking}

In this section, we prove that the difference between the broadcast domination number and the multipacking number of the cactus and AT-free graphs can be arbitrarily large.  
 Moreover, we make a connection with the \emph{fractional} versions of the two concepts dominating broadcast and multipacking for cactus and AT-free graphs.

\begin{figure}[ht]
    \centering

\includegraphics[height=3cm]{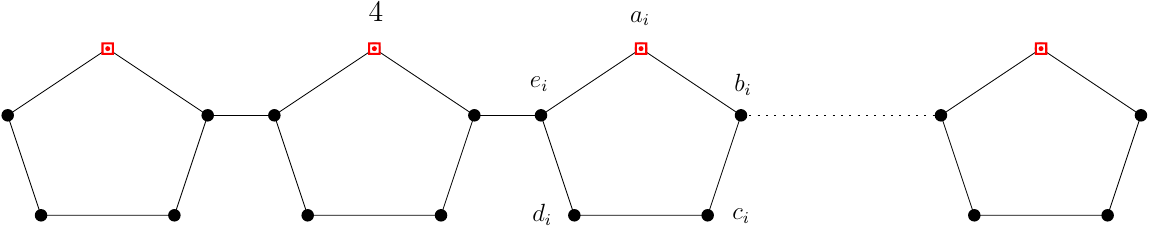}
    \caption{The $G_k$ graph with $\gamma_b(G_k)=4k$ and $\MP(G_k)=3k$. The set $\{a_i:1\leq i\leq 3k\}$ is a maximum multipacking of $G_k$.} 
    \label{fig:pentagon}
\end{figure}

To prove that the difference $ \gamma_{b}(G) -  \MP(G) $ can be arbitrarily large, we construct the graph $G_k$ as follows. Let $A_i=(a_i,b_i,c_i,d_i,e_i,a_i)$ be a 5-cycle  for each  $i=1,2,\dots,3k$. We form $G_k $ by joining $b_i$ to $e_{i+1}$ for each $i=1,2,\dots,3k-1$ (See Fig. \ref{fig:pentagon}). This  graph is a  cactus graph (and AT-free graph). We show that $\MP(G_k)=3k$ and $\gamma_b(G_k)=4k$.

\begin{lemma} \label{lem:mpGk=3k} $\MP(G_{k})=3k$, for each positive integer $k$.
\end{lemma}

\begin{proof} The path $P=(e_1,a_1,b_1,e_2,a_2,b_2,\dots,e_{3k}a_{3k}b_{3k})$ is a diametral path of $G_k$ (Fig.\ref{fig:pentagon}). This implies $P$ is an isometric path of $G_k$ having the length  $l(P)=3.3k-1$. By Lemma \ref{lem:isometric_path}, every third vertex on this path form a multipacking of size $\big\lceil\frac{3.3k-1}{3}\big\rceil=3k$. Therefore, $\MP(G_k)\geq 3k$. Note that, diameter of $A_i$ is $2$ for each $i$. Therefore, any multipacking of $G_k$ can contain at most one vertex of $A_i$ for each $i$. So, $\MP(G_k)\leq 3k$. Hence $\MP(G_k)= 3k$.
\end{proof}




\noindent\textbf{Fractional multipacking: }
R. C. Brewster and  L. Duchesne \cite{brewster2013broadcast} introduced fractional multipacking in 2013 (also see \cite{teshima2014multipackings}).  Suppose $G$ is a graph with $V(G)=\{v_1,v_2,v_3,\dots,v_n\}$ and $w:V(G)\rightarrow [0,\infty)$  is a function. So, $w(v)$ is a weight on a vertex $v\in V(G)$. Let $w(S)=\sum_{u\in S}w(u)$ where $S\subseteq V(G)$.  We say   $w$ is a \textit{fractional multipacking} of $G$,  if $w( N_r[v])\leq r$ for each vertex $ v \in V(G) $ and for every integer $ r \geq 1 $. The \textit{fractional multipacking number} of $ G $ is the  value $\displaystyle \max_w w(V(G)) $ where $w$ is any fractional multipacking and it
	is denoted by $ \MP_f(G) $. A \textit{maximum fractional multipacking} is a fractional multipacking $w$  of a graph $ G $ such that	$ w(V(G))=\MP_f(G)$. If $w$ is a fractional multipacking, we define   a vector $y$ with the entries $y_j=w(v_j)$.  So,  \begin{center}
	    $\MP_f(G)=\max \{y.\mathbf{1} :  yA\leq c, y_{j}\geq 0\}.$
	\end{center} So, this is a  linear program which is the dual of the linear program   $\min \{c.x :  Ax\geq \mathbf{1}, x_{i,k}\geq 0\}$. Let,  
	    $\gamma_{b,f}(G)=\min \{c.x : Ax\geq \mathbf{1}, x_{i,k}\geq 0\}.$
	
 Using the strong duality theorem for linear programming, we can say that \begin{center}
     $\MP(G)\leq \MP_f(G)= \gamma_{b,f}(G)\leq \gamma_{b}(G).$
 \end{center}


\begin{figure}[ht]
    \centering
    \includegraphics[height=3cm]{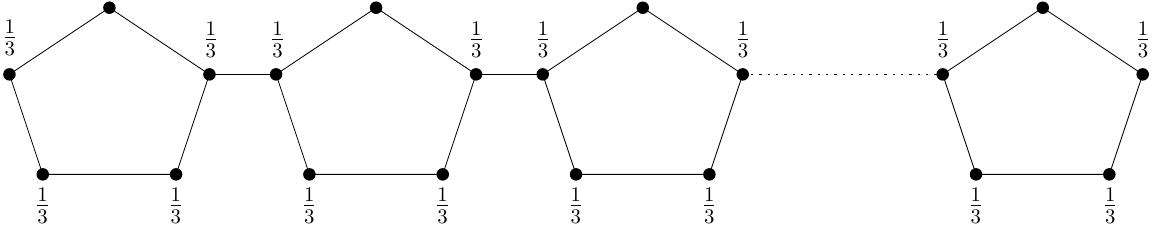}
    \caption{The $G_k$ graph with $\MP_f(G_k)=4k$.}
    \label{fig:pentagon_frac}
\end{figure}

\begin{lemma}  \label{lem:mpfGk=gammabGk=4k}
If $k$ is a positive integer, then $\MP_f(G_{k})=\gamma_b(G_{k})= 4k$.
\end{lemma}

\begin{proof}  
We define a function $w: V(G_k)\rightarrow [0,\infty)$  where $w(b_{i})=w(c_{i})=w(d_{i})=w(e_{i})=\frac{1}{3}$  for each $i=1,2,3,\dots,3k$ (Fig. \ref{fig:pentagon_frac}).  So, $w(G_k)=4k$. 
 We want to show that $w$ is a fractional multipacking of $G_{k}$. So, we have to prove that $w(N_r[v])\leq r$ for each vertex $ v \in V(G_{k}) $ and for every integer $ r \geq 1 $. We prove this statement using induction on $r$.	It can be checked that $w(N_r[v])\leq r$ for each vertex $ v \in V(G_{k}) $ and for each  $ r \in \{1,2,3,4\} $. Now assume that the statement is true for $r=s$, we want to prove that it is true for $r=s+4$. Observe that, $w(N_{s+4}[v]\setminus N_{s}[v])\leq 4$, $\forall v\in V(G_{k})$. Therefore,  $w(N_{s+4}[v])\leq w(N_{s}[v])+4\leq s+4$. So, the statement is true. So, $w$ is a fractional multipacking of $G_{k}$. Therefore, $\MP_f(G_{k})\geq 4k$.

Define a broadcast ${f}$ on $G_k$ as ${f}(v)=
    \begin{cases}
        4 & \text{if } v=a_i  \text{ and } i\equiv 2 \text{ (mod $3$)}  \\
        0 & \text{otherwise }
    \end{cases}$.\\
Here  ${f}$ is an efficient dominating broadcast and $\sum_{v\in V(G_k)}{f}(v)=4k$. So, $\gamma_b(G_k)\leq 4k$, for all $ k\in \mathbb{N}$. So, by the strong duality theorem  we have  $4k\leq \MP_f(G_k)= \gamma_{b,f}(G_k)\leq \gamma_{b}(G_k)\leq 4k$. Therefore, $\MP_f(G_{k})=\gamma_b(G_{k})= 4k$.  
\end{proof}

Lemma \ref{lem:mpGk=3k}  and  Lemma \ref{lem:mpfGk=gammabGk=4k} imply the following results.

\multipackingbroadcastgape*


\gammabDIFFGmpG*





\begin{corollary} \label{cor:mpfG-mpG}  The difference $\MP_f(G)-\MP(G)$ can be arbitrarily large for cactus graphs (and for AT-free graphs).
\end{corollary}



\section{$1/2$ - hyperbolic graphs}\label{sec:1/2-Hyperbolic graphs}

In this section, we show that the family of graphs $G_k$ (Fig. \ref{fig:pentagon}) is $\frac{1}{2}$-hyperbolic using a characterization of $\frac{1}{2}$-hyperbolic graphs. Using this fact, we  show that the difference $ \gamma_{b}(G) -  \MP(G) $ can be arbitrarily large for $\frac{1}{2}$-hyperbolic graphs. 
If the length of a shortest path $P$ between two vertices $x$ and $y$ of a cycle $C$ of $G$ is  smaller than the distance between $x$ and $y$ measured along $C$, then $P$ is called a \textit{bridge} of $C$. In $G$, a cycle $C$ is called \textit{well-bridged}  if for any vertex $x \in C$ there exists a bridge from $x$ to some vertex of $C$ or if the two neighbors of $x$ from $C$ are adjacent.

\begin{figure}[ht]
    \centering
    \includegraphics[height=5.5cm]{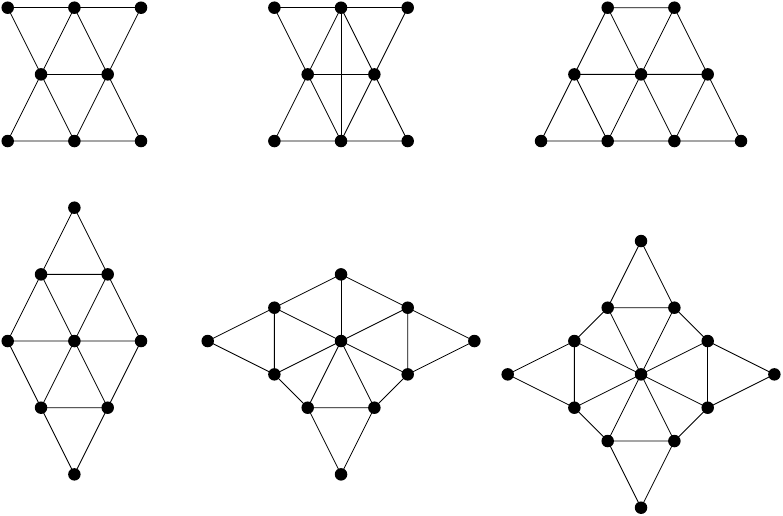}
    \caption{Forbidden isometric subgraphs for $\frac{1}{2}$-hyperbolic graphs} 
    \label{fig:hyperbolic_forbidden_graphs}
\end{figure}

\begin{theorem}[\cite{bandelt20031}] \label{thm:1/2-Hyperbolic_graphs_classification}
    A graph $G$ is $\frac{1}{2}$-hyperbolic if and only if all cycles $C_n$, $n \neq 5$, of $G$ are well-bridged and none of the graphs in Figure \ref{fig:hyperbolic_forbidden_graphs} occur as isometric subgraphs of $G$.
\end{theorem}

Theorem \ref{thm:1/2-Hyperbolic_graphs_classification} yields the following.

\begin{lemma}\label{lem:pentagon_half_hyperbolic}
    The family of graphs $G_k$ (Fig. \ref{fig:pentagon}) is $\frac{1}{2}$-hyperbolic.
    
\end{lemma}

Lemma \ref{lem:mpGk=3k}, Lemma \ref{lem:mpfGk=gammabGk=4k} and Lemma \ref{lem:pentagon_half_hyperbolic} imply the following theorem.

\hypermultipackingbroadcastgape*

\hypergammabDIFFGmpG*


From Theorem \ref{thm:1/2-Hyperbolic_graphs} and Lemma \ref{lem:mpfGk=gammabGk=4k}, we have the following.

\begin{corollary} \label{cor:hyper_mpfG-mpG}  The difference $\MP_f(G)-\MP(G)$ can be arbitrarily large for $\frac{1}{2}$-hyperbolic graphs.
\end{corollary}

\section{Bound of the ratio between broadcast domination and multipacking numbers of connected chordal graphs}\label{sec:chordal graphs}

In this section, we improve the lower bound of the expression $\lim_{\MP(G)\to \infty}$ $\sup\{\gamma_{b}(G)/\MP(G)\}$ for connected chordal graphs to $4/3$ where the previous lower bound was $10/9$  \cite{das2023relation}. To show this, we construct the graph $F_k$ as follows. Let $A_i$ be a 
 graph having the vertex set $\{a_i,b_i,c_i,d_i,e_i,g_i\}$ where  $(g_i,a_i,b_i,c_i,d_i,e_i,g_i)$ is a 6-cycle and $(a_i,c_i,e_i)$ is a 3-cycle,  for each  $i=1,2,\dots,3k$. We form $F_k $ by joining $b_i$ to $g_{i+1}$ for each $i=1,2,\dots,3k-1$ (See Fig. \ref{fig:pentagon}). This  graph is a  connected chordal graph. We show that $\MP(F_k)=3k$ and $\gamma_b(F_k)=4k$.

 \begin{figure}[ht]
    \centering

\includegraphics[height=3.1cm]{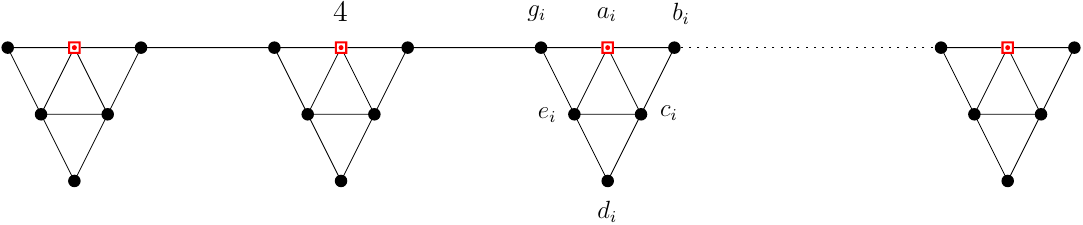}
    \caption{The $F_k$ graph with $\gamma_b(F_k)=4k$ and $\MP(F_k)=3k$. The set $\{a_i:1\leq i\leq 3k\}$ is a maximum multipacking of $F_k$.} 
    \label{fig:chordal}
\end{figure}

\begin{lemma} \label{lem:chordal_mpFk=3k} $\MP(F_{k})=3k$, for each positive integer $k$.
\end{lemma}

\begin{proof} The path $P=(g_1,a_1,b_1,g_2,a_2,b_2,\dots,g_{3k}a_{3k}b_{3k})$ is a diametral path of $F_k$ (Fig.\ref{fig:chordal}). This implies $P$ is an isometric path of $F_k$ having the length  $l(P)=3.3k-1$. By Lemma \ref{lem:isometric_path}, every third vertex on this path form a multipacking of size $\big\lceil\frac{3.3k-1}{3}\big\rceil=3k$. Therefore, $\MP(F_k)\geq 3k$. Note that, diameter of $A_i$ is $2$ for each $i$. Therefore, any multipacking of $F_k$ can contain at most one vertex of $A_i$ for each $i$. So, $\MP(F_k)\leq 3k$. Hence $\MP(F_k)= 3k$.
\end{proof}

\begin{figure}[ht]
    \centering
    \includegraphics[height=3.1cm]{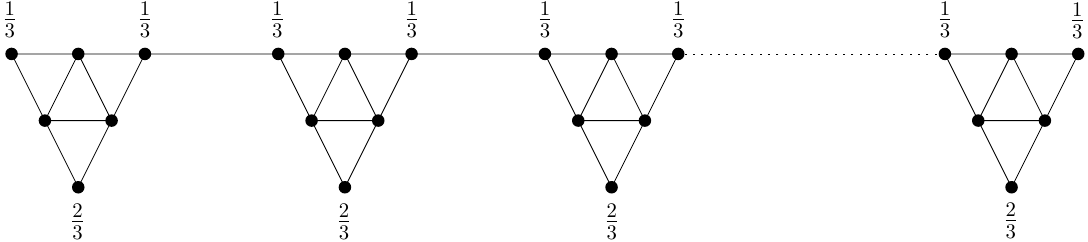}
    \caption{The $F_k$ graph with $\MP_f(F_k)=4k$.}
    \label{fig:chordal_frac}
\end{figure}

\begin{lemma}  \label{lem:chordal_mpfFk=gammabFk=4k}
If $k$ is a positive integer, then $\MP_f(F_{k})=\gamma_b(F_{k})= 4k$.
\end{lemma}

\begin{proof}  
We define a function $w: V(F_k)\rightarrow [0,\infty)$  where $w(g_{i})=w(b_{i})=\frac{1}{3}$ and $w(d_{i})=\frac{2}{3}$   for each $i=1,2,3,\dots,3k$ (Fig. \ref{fig:chordal_frac}).  So, $w(F_k)=4k$. 
 We want to show that $w$ is a fractional multipacking of $F_{k}$. So, we have to prove that $w(N_r[v])\leq r$ for each vertex $ v \in V(F_{k}) $ and for every integer $ r \geq 1 $. We prove this statement using induction on $r$.	It can be checked that $w(N_r[v])\leq r$ for each vertex $ v \in V(F_{k}) $ and for each  $ r \in \{1,2,3,4\} $. Now assume that the statement is true for $r=s$, we want to prove that it is true for $r=s+4$. Observe that, $w(N_{s+4}[v]\setminus N_{s}[v])\leq 4$, $\forall v\in V(F_{k})$. Therefore,  $w(N_{s+4}[v])\leq w(N_{s}[v])+4\leq s+4$. So, the statement is true. So, $w$ is a fractional multipacking of $F_{k}$. Therefore, $\MP_f(F_{k})\geq 4k$.

Define a broadcast ${f}$ on $F_k$ as ${f}(v)=
    \begin{cases}
        4 & \text{if } v=a_i  \text{ and } i\equiv 2 \text{ (mod $3$)}  \\
        0 & \text{otherwise }
    \end{cases}$.\\
Here  ${f}$ is an efficient dominating broadcast and $\sum_{v\in V(F_k)}{f}(v)=4k$. So, $\gamma_b(F_k)\leq 4k$, for all $ k\in \mathbb{N}$. So, by the strong duality theorem  we have  $4k\leq \MP_f(F_k)= \gamma_{b,f}(F_k)\leq \gamma_{b}(F_k)\leq 4k$. Therefore, $\MP_f(F_{k})=\gamma_b(F_{k})= 4k$.  
\end{proof}

Lemma \ref{lem:chordal_mpFk=3k}  and  Lemma \ref{lem:chordal_mpfFk=gammabFk=4k} imply the following results.

\chordalmultipackingbroadcastgape*

Therefore, the range of the expression $\lim_{\MP(G)\to \infty}$ $\sup\{\gamma_{b}(G)/\MP(G)\}$ which was previously $[10/9,3/2]$  \cite{das2023relation} is improved to $[4/3,3/2]$.

\begin{restatable}{corollary}{chordalgammabGBYmpG} \label{cor:chordal_gammabG/mpG} For connected chordal graphs $G$, \text{ }
$\displaystyle\frac{4}{3}\leq\lim_{\MP(G)\to \infty}\sup\Bigg\{\frac{\gamma_{b}(G)}{\MP(G)}\Bigg\}\leq \frac{3}{2}.$
\end{restatable}

\section{Conclusion}\label{sec:conclusion}

We have shown that the  bound $\gamma_b(G)\leq 2\MP(G)+3$ for general graphs $G$ can be improved to $\gamma_b(G)\leq \frac{3}{2}\MP(G)+\frac{11}{2}$ for cactus graphs. Additionally, we have given a $(\frac{3}{2} + o(1))$-factor approximation algorithm for the multipacking problem on cactus graphs. A natural direction for future research is to investigate whether a polynomial-time algorithm exists for finding a maximum multipacking on cactus graph.

It remains an interesting open problem to determine the best possible value of the expression $\lim_{\MP(G)\to \infty}$ $\sup\{\gamma_{b}(G)/\MP(G)\}$ for general connected graphs. Furthermore, improving the bounds for the expression in subclasses such as connected chordal graphs and cactus graphs presents an interesting avenue for further study. Extending this investigation to other important graph classes could also yield valuable insights into the relationship between broadcast domination and multipacking numbers.

\bibliographystyle{elsarticle-num}
\bibliography{ref.bib}











\end{document}